\documentclass[a4paper,11pt]{article}
\usepackage{amsmath}
\usepackage{amssymb}
\usepackage{cite}
\usepackage{epsfig}
\usepackage{cancel}
\usepackage{feynmp}
\usepackage{xspace}
\usepackage{flafter}

\def\spa#1.#2{\left\langle#1\,#2\right\rangle}
\def\spb#1.#2{\left[#1\,#2\right]}
\def\spaa#1.#2.#3{\langle\mskip-1mu{#1}
                  | #2 | {#3}\mskip-1mu\rangle}
\def\spbb#1.#2.#3{[\mskip-1mu{#1}
                  | #2 | {#3}\mskip-1mu]}
\def\spab#1.#2.#3{\langle\mskip-1mu{#1}
                  | #2 | {#3}\mskip-1mu\rangle}
\def\spba#1.#2.#3{\langle\mskip-1mu{#1}^+
                  | #2 | {#3}^+\mskip-1mu\rangle}
\def\spav#1.#2.#3{\|\mskip-1mu{#1}
                  | #2 | {#3}\mskip-1mu\|^2}
\def\jc#1.#2.#3{j^{#1}_{#2#3}}
\def\blfootnote{\xdef\@thefnmark{}\@footnotetext} 
\newcommand{\Ca}{\ensuremath{C_{\!A}}\xspace}
\newcommand{\Cf}{\ensuremath{C_{\!F}}\xspace}
\newcommand{\nf}{\ensuremath{n_{\!f}}\xspace}
\newcommand{\Nc}{\ensuremath{N_{\!C}}\xspace}
\newcommand{\gs}{\ensuremath{g}}
\newcommand{\as}{\ensuremath{\alpha_s}\xspace}

\date{January 27, 2011}
\title{
\begin{normalsize}
\begin{flushright}
CP3-Origins-2011-02\\
Edinburgh 2011/03
\end{flushright}
\end{normalsize}
\vspace*{1cm}
Multiple Jets at the LHC with High Energy Jets}
\author{Jeppe~R.~Andersen$^{a}$,
  Jennifer~M.~Smillie$^{b}$\\\mbox{}\\$^a$ {CP}$^{ \bf 3}${-Origins}, 
  Campusvej 55, DK-5230 Odense M, Denmark. \\$^b$ School of Physics and Astronomy, University of Edinburgh,\\
         Mayfield Road, Edinburgh EH9 3JZ, UK.}

\begin{document}
\maketitle
\begin{abstract}
  We present a flexible Monte Carlo implementation of the perturbative
  framework of \emph{High Energy Jets}, describing multi-jet events at hadron
  colliders. The description includes a resummation which ensures leading
  logarithmic accuracy for large invariant mass between jets, and is matched
  to tree-level accuracy for multiplicities up to 4 jets. The resummation
  includes all-order hard corrections, which become important for increasing
  centre-of-mass energy of the hadronic collision.

  We discuss observables relevant for confronting the perturbative framework
  with 7~TeV data from the LHC, and the impact of the perturbative
  corrections on several dijet and trijet observables which are relevant in
  the search for new physics.
\end{abstract}
\newpage
\tableofcontents
\section{Introduction}
\label{sec:introduction}
The cross-section at the LHC for particles charged under QCD will generally
be larger than that for colourless particles, and so many of the discovery
channels used in the search for new physics involve the detection of hard,
hadronic jets. The large mass hierarchy between any (often heavy, in order to
avoid existing exclusion limits) new particle produced and those of the decay
products often implies that many jets will be produced in the decay of a new
state. The finger prints of any such new physics will, however, have to be
found amongst a large contribution to the same signature from multi-jet
processes within the Standard Model. Therefore, a detailed understanding of
the Standard Model processes will assist in the search for new
physics. Examples of Standard Model processes acting as background to many
searches for new physics are e.g.~$W,Z+$jets (especially with 3,4 jets or
more).

However, even the nature of some Standard Model processes is best studied in
events with multiple jets. For example, the $CP$-structure of the induced
Higgs boson couplings to gluons through a top-loop could be measured by a
study of the azimuthal angle between the two jets in events with a Higgs
boson in association with dijets\cite{Klamke:2007cu,Andersen:2010zx}.

In both examples, hard radiative corrections will be sizeable at the LHC, by
which we mean that the exclusive $(n+1)$-jet rate is a significant component
of the inclusive $n$-jet rate. And more so in many of the regions of
interest for searches of new physics. Therefore, a tree-level description of
the inclusive $n$-jet process will be unsatisfactory for the involved
analyses beyond just a measurement of the cross section. 

The reason for the
increased importance in many situations of hard, perturbative corrections at
the LHC over the situation at previously, lower energy colliders is very
simple. Two effects act to suppress hard corrections: the increasing powers
of the perturbative coupling, and the increase in the light-cone
momentum fraction of the partons extracted from the proton beyond that
necessary for the final state without the additional hard jet. The
suppression from this last kinematic effect is caused by the decrease in the
parton density functions (pdf) as the light-cone momentum fraction $x$ is
increased. However, for processes with at least two particles in the final
state, there is a fine trade-off between the suppression from the pdf and the
increasing phase space for additional emission in-between the most
forward/backward hard jet (even when this additional emission is hard in transverse
momentum), as the rapidity span between the two most forward/backward jets is
increased. At 
previous, lower-energy colliders, this balance was tipped more towards a
suppression than will be the case at the LHC. 

At previous colliders, the
``significant'' rapidity separation of the two objects, which is necessary
for the opening of phase space for additional radiation, would already bring
the light-cone momentum fractions into the region of extremely fast falling
pdfs as $x\to1$, thus effectively vetoing additional emissions. However, the
situation is different for the LHC processes discussed above, since in the
case of e.g.~$W$-boson production with at least 3 jets, two jets will naturally be
produced with a size-able separation in rapidity\cite{Binoth:2010ra}. In this
case, there is only a small suppression for additional (especially central)
radiation, even when the additional jets have a sizeable transverse momentum.
This holds true also for other processes, provided the hard scattering
amplitude has a mechanism for effectively radiating into the rapidity
span. This is the case when colour is exchanged between the particles either
side of the span, whereas a colour-singlet exchange leads to less radiation
in the span\cite{Dokshitzer:1987nc,Dokshitzer:1991he}.

While a fixed order (e.g.~LO or NLO) calculation may be adequate for the
description of sufficiently inclusive quantities like the total inclusive
cross section, the question is to what extent a given theoretical
description allows for the radiation into the phase space which becomes
available with the increase in partonic centre of mass energy --- and how
important the description of this radiation is for a given observable. The
current paper discusses these problems, and presents results obtained in a
recently proposed all-order perturbative framework.

It is clear that a NLO calculation allows for just one, also hard, additional
emission above the minimum number of jets required in the analysis. The
all-order description of a parton shower, on the other hand, captures the
soft and collinear emissions, but will underestimate the amount of hard
radiation. This deficiency can be repaired order-by-order through a
CKKW-L-style matching\cite{Catani:2001cc,Lonnblad:1992tz,Mangano:2001xp}, or
to full NLO accuracy\cite{Frixione:2002ik,Nason:2004rx} for low
multiplicities. In both cases, the deficiency of the parton shower in
describing hard radiation is repaired by the use of full tree-level matrix
elements. The maximum multiplicity applied in the tree-level matching is
limited by the time for evaluation of the tree-level matrix elements.  Since
in a CKKW-L-style analysis the matching scale should be chosen somewhat
smaller than the transverse scale required in the definition of jets to avoid
matching artefacts, the matching procedure will run out of available matrix
elements at a lower multiplicity than the maximum for which the LO process
has been calculated. This can be viewed simply as a result of the attempt
within the matching procedure to describe not just the total rate, but also
the final state configuration.

The framework of \emph{High Energy Jets} (HEJ)\cite{Andersen:2009nu,Andersen:2009he}
provides an all-order description of processes with more than two hard jets,
based on an approximation which captures the hard, wide-angle emissions
missed in a shower-approach based on soft and collinear splitting
functions. HEJ does not try to redo the job of the shower, but focuses
specifically on the part \emph{not} done by a parton shower. Work is in
progress to combine the description of HEJ with a parton
shower\cite{Andersen:soon}; the most important component of the matching
between HEJ and a parton shower is the avoidance of double counting of soft
radiation, which is treated to all orders in both descriptions. 

The formalism of HEJ is inspired by that
underlying\cite{Fadin:1975cb,Kuraev:1976ge,Kuraev:1977fs} the BFKL equation\cite{Balitsky:1978ic}, and
as such, an approximation for both real and virtual corrections is obtained
to all orders, obviously with all IR divergences cancelling between the two
contributions. Differently to the BFKL approach, however, HEJ applies an
approximation only to the partonic scattering amplitudes, and not the phase space
integration, which is performed for each explicit multiplicity. In this
respect, HEJ resembles a parton shower formulation of an all-order
summation. Furthermore, by applying the approximations at the level of the
scattering amplitude $\mathcal{M}$ (and not $|\mathcal{M}|^2$), it is
possible to supplement\cite{Andersen:2009nu,Andersen:2009he} the
approximations with the requirement of e.g.~gauge invariance, and thereby
obtain a formalism, which reproduces more accurately the fixed order
perturbative results when checked order by order, while simultaneously being
sufficiently simple that all-order results can be explicitly obtained.

In the current paper, we develop further the formalism of High Energy Jets by
matching to fixed order results and include some sub-leading
corrections. Furthermore, we demonstrate the application of HEJ to the
production of at least two and at least three jets. 

The structure of the paper is as follows: in Section~\ref{sec:high-energy-jets}
we briefly review the formalism within \emph{High Energy Jets}, which allows
approximate all-order results to be
obtained\cite{Andersen:2009nu,Andersen:2009he}. In Section~\ref{sec:matching} we
describe the matching of these amplitudes to full, high-multiplicity
tree-level results. In Section~\ref{sec:running-coupling} we include some
sub-leading corrections, which stabilise the dependence on the scale choice\cite{Andersen:2003an,Andersen:2003wy}. In Section~\ref{sec:dijet-event-results} we present
results for dijet production obtained with the full formalism of \emph{High
  Energy Jets}, and discuss observables and distributions for which the
higher-order corrections are particularly important in order to obtain a
perturbatively stable description. These can lead to a direct experimental
test of the importance of the correct perturbative description.

The all-order results presented in this paper are obtained using the
implementation of the formalism of \emph{High Energy Jets} in a fully
flexible parton-level Monte Carlo generator, which can be downloaded at
\textsc{http://cern.ch/hej}.


\section{All Orders with High Energy Jets}
\label{sec:high-energy-jets}
The all-order perturbative framework of \emph{High Energy Jets} (\emph{HEJ})
initiated in
Ref.\cite{Andersen:2008ue,Andersen:2008gc,Andersen:2009nu,Andersen:2009he} is
addressing some of the short-comings in the description of multiple hard,
perturbative corrections in both the (low) fixed-order and in the parton
shower formulation. The perturbative description obtained with \emph{HEJ}
reproduces the correct, all-order, full QCD limit for both real and virtual
corrections to the hard perturbative matrix element for the hard, wide-angle
emissions which underpin the perturbative description of the formation of
additional jets. The central parts of the formalism were presented in
Ref.\cite{Andersen:2009nu,Andersen:2009he} and discussed further in
Ref.\cite{Andersen:2010ch,Andersen:2010ih}. In this section, we will give just a brief
overview of the formalism on which the approximations are based; the next
section will then discuss how to incorporate matching corrections to full,
high multiplicity tree-level accuracy.

\boldmath
\subsection{Dominance of the $t$-channel poles, and current-current  scatting}
\label{sec:dominance-t-channel}
\unboldmath 
In the standard parton shower formalism, the physical picture is one of
successive branchings off $s$-channel propagators, governed by the
DGLAP splitting
functions~\cite{Gribov:1972ri,Lipatov:1974qm,Altarelli:1977zs,Dokshitzer:1977sg}. Such a
framework can sum to leading 
logarithmic accuracy, and to all orders the behaviour dictated by the \emph{soft} and
\emph{collinear} $s$-channel singularities arising in the perturbative corrections to a given
scattering amplitude. It describes correctly emissions with small invariant
mass to the hard scattering amplitude.

The limit of pure $N$-jet amplitudes for \emph{large} invariant mass between
each parton of similar transverse momentum is described by the
FKL-amplitudes\cite{Fadin:1975cb,Kuraev:1976ge}, which are at the foundation
of the BFKL framework\cite{Balitsky:1978ic}. The physical picture arising
from the FKL amplitudes is one of effective vertices connected by $t$-channel
propagators. The reduction of the formalism to the two-dimensional BFKL
integral equation relies on many kinematical approximations, which are
extended to all of phase space. Using an explicit (or so-called iterative)
solution to the BFKL equation\cite{Schmidt:1996fg,Orr:1997im,Andersen:2001kt}, it is however
straightforward to show that despite the logarithmic accuracy (in $\hat s
/\hat t$), the perturbative expansion of the (B)FKL solution does not give a
satisfactory description of the results obtained order by order with the true
perturbative series from QCD\cite{Andersen:2008gc}.

\emph{High Energy Jets}\cite{Andersen:2009nu,Andersen:2009he} inherits the
idea of effective vertices connected by $t$-channel currents in order to
reproduce the correct limit of $N$-jet amplitudes, but goes beyond
controlling just the logarithmic accuracy like the FKL formalism. The kinematic
building blocks of the FKL formalism depend on transverse momenta only, as a
result of the kinematic limits applied in order to separate the amplitude
into effective vertices separated by $t$-channel
exchanges\cite{Fadin:2006bj}. In the following, we will discuss how to obtain
a better approximation for the $t$-channel singularities.

The $2\to2$ scattering $qQ\to qQ$ obviously proceeds through just a
$t$-channel exchange of the gluon current generated by a quark. A good
formalism for the description the $t$-channel poles should get at least this
very simple process exact. The colour and helicity averaged and summed square
of this simple scattering amplitude is given by
\begin{align}
  \label{eq:ME2qQtoqQ}
  |\overline{\mathcal{M}}^{\mathrm{tree}}_{qQ\to qQ}|^2=g^4\ \frac 4 9\ \frac{s^2+u^2}{t^2}.
\end{align}
Despite its simplicity, this amplitude can already be used to illustrate the
problem of the approximations made in the standard BFKL procedure. The limit
of \emph{Multi-Regge-Kinematics} is defined for the scattering process
$p_A,p_B\to p_1,\ldots, p_n$ in terms of transverse momenta and rapidities
$y=\ln\left(\frac{E+p_z}{E-p_z}\right)$ as the following conditions
\begin{align}
  \label{eq:MRKlimit}
    \forall i\in \{2,\ldots, n-1\}: y_{i-1}\gg y_i \gg y_{i+1}, \quad
    \forall i,j: |p_{i\perp}|\approx |p_{j\perp}|,
\end{align}
or alternatively
\begin{align}
  \label{eq:MRKlimit2}
    \forall i,j: |p_{i\perp}|\approx |p_{j\perp}|, \quad s_{ij}\to \infty,
\end{align}
where $s_{ij}=2\ p_i.p_j$ and $s=2\ p_A.p_B$. For the $2\to2$ process, the MRK
limit of the Mandelstam variables is given by $t\to-k_\perp^2$,
$s\approx-u\to\infty$. The effective approximation applied in the BFKL
formalism (both at LL and NLL) for the $2\to2$ process is
$|\overline{\mathcal{M}}^{\mathrm{BFKL,Tree}}_{qQ\to qQ}|^2=g^4\ \frac 8 9\
s^2/(k_\perp^2)^2$. However, for much of the kinematics relevant at the LHC, $t$
and $-k_\perp^2$ differ by at least an order of magnitude, and $s$ and $u$
differ significantly, leading to a gross overestimation of the cross section,
if the BFKL approximation is applied.

In Eq.~\eqref{eq:ME2qQtoqQ}, the $s^2$-component arises from scattering of
quarks of the same helicities (e.g.~$q^-Q^-\to q^-Q^-$), whereas the
$u^2$-component arises from the scatting of unlike helicities
(e.g.~$q^-Q^+\to q^-Q^+$). Since this difference is important in obtaining
sufficient accuracy, \emph{HEJ} is based on the calculation of scattering
processes at the amplitude level (as opposed to the square of the amplitude),
and the sum over helicities is performed explicitly. For the $qQ$-process then,
the obvious choice of formalism is that of current-current scattering. 

In the spinor notation for the quark currents (see Ref.\cite{Andersen:2009nu} for
details), $j_{a1}^{-\mu}=\bar{u}_1^- \gamma^\mu u_a^-$ is written as $\spab1.\mu.a$, and then
the (colour and coupling stripped) matrix element for the process $q^-_{p_a}Q^-_{p_b}\to
q^-_{p_1}Q^-_{p_2}$ reads
\begin{align}
  \label{eq:MqQ}
  M_{q^-Q^-\to q^-Q^-} &=\ \spab1.\mu.a \frac{g^{\mu\nu}}{t}\spab 2.\nu.b.
\end{align}
While it is possible to shorten this expression by use of the Fierz identity, we
choose to keep the formulation in terms of currents, as this will prove
useful for the generalisation to other processes, including $W,H,Z$+jets.

Let us denote the spinor string (for helicities $h_a,h_1,h_b,h_2$ of the
quarks) appearing in the amplitude as
\begin{align}
  \label{eq:spinorstring}
  S^{h_ah_b\to h_1h_2}_{qQ\to qQ}=\langle 1\ h_1|\mu|a\ h_a\rangle\
  g^{\mu\nu}\ \langle 2\ h_2|\nu| b\ h_b\rangle.
\end{align}
This complex number can be calculated using any explicit representation for
the spinors (see e.g.~Ref.\cite{Andersen:2009nu,Andersen:2009he}), and we
will denote the sum over helicities of the absolute square of this number by
\begin{align}
  \label{eq:spinorsqsum}
  \left\|S_{qQ\to qQ}\right\|^2=\sum_{h_a,h_a,h_b,h_2} \left |S^{h_ah_b\to h_1h_2}_{qQ\to qQ}\right|^2.
\end{align}
Of course in this case non-zero contributions arise only when $h_a=h_1$ and $h_b=h_2$.

The colour and helicity summed and averaged matrix element for the scattering
process $qQ\to qQ$ is then
\begin{align}
  \begin{split}
    \label{eq:MqQqQchsa}
    \overline{\left|\mathcal{M}_{qQ\to qQ}\right|}^2\ =\ &\frac 1 {4\
      (\Nc^2-1)}\ \left\|S_{qQ\to qQ}\right\|^2\\
    &\cdot\ \left(g^2\ \Cf\ \frac 1 {t_1}\right)\\
    &\cdot\ \left(g^2\ \Cf\ \frac 1 {t_2}\right).
  \end{split}
\end{align}
with $t_1=(p_a-p_1)^2$ and $t_2=(-p_b+p_2)^2$ (obviously $t_1=t_2$ in this
case of a $2\to2$-process), which equals Eq.~\eqref{eq:ME2qQtoqQ}.

The point of this tour de force through the simple formalism of
$qQ$-scattering is that using this formalism, the amplitudes for $qg$-scattering
can be recast in a very similar form. In fact, a careful
analysis\cite{Andersen:2009he} of the helicity structure in $qg\to
qg$-scattering reveals that all the amplitudes where the helicity of the
gluon is unchanged\footnote{All helicity-flip amplitudes are systematically
  suppressed by a factor of $\hat s$.} factorise again into two currents
contracted over a $t$-channel pole. For example, the fully colour-dressed
scattering amplitude for the process $q^-(p_a) + g^+(p_b) \to q^-(p_1) +
g^+(p_2)$ equals\cite{Andersen:2009he}
\begin{equation}
  \label{eq:qmgmsum}
  \mathcal{M}_{q^-g^+\to q^-g^+}={-ig^2}\  \frac{p_{2\perp}^*}{|p_{2\perp}|} \left(
    t^2_{ea}t^b_{1e}
    \sqrt{\frac{p_b^-}{p_2^-}} - t^b_{ea} t^2_{1e} \sqrt{\frac{p_2^-}{p_b^-}} \right)
  \spab1.\mu.a\  \frac{g^{\mu\nu}}{t}\ \spab{b}.\nu.2,
\end{equation}
with $p_{\perp}=p_x+i\ p_y$, $p^-=E-p_z$. We
have taken the negative $z$-direction to be that of the incoming gluon, without loss of generality. We
immediately recognise the kinematic structure (in terms of currents) of
$q^-Q^+$-scattering, multiplied by a momentum-dependent
colour factor. The colour summed and averaged scattering matrix element is
\begin{align}
  \begin{split}
    \label{eq:sqcol}
    |\mathcal{M}_{q^-g^+\to q^- g^+}|^2\ &=  \frac 1 {\Nc^2-1}\ | \spab{b}.\rho.2 \spab1.\rho.a |^2\\
    &\cdot \left( g^2\ \Cf\ \frac 1 {t_1}\right)\\
    &\cdot \left( g^2\  \left[ \frac 1 2\ 
       \frac{1+z^2}{z}\ \left(C_A -\frac 1 {C_A}\right)+
       \frac{1}{C_A}\right]\ \frac 1 {t_2}\right),
  \end{split}
\end{align}
where $z=p_2^-/p_b^-$ (and again $t_1=(p_a-p_1)^2=(p_b-p_2)^2=t_2$). This has
a striking similarity to the amplitude for $qQ$-scattering (see
Eq.~\eqref{eq:MqQqQchsa}). In fact, it differs only by the slightly more
complicated colour factor in square brackets, which replace the \Cf in the case
of quark scattering. In the MRK limit $p_2^-\to
p_b^-$ this tends to \Ca, and the $qg$ scattering matrix element is equal to
the one for $qQ$ rescaled by $\Ca/\Cf$, just as
expected\cite{Combridge:1984jn}. Eq.~\eqref{eq:sqcol} is, however, the exact
result, and the square bracket is strictly larger than \Ca, and uniformly
decreasing for increasing $z$. Small $z$ here reflects a large change in
light-cone momentum for the gluon, and unsurprisingly a strong acceleration is
reflected in an effectively stronger interaction (though this is unrelated to
the higher order perturbative effect of the running of the coupling).

The perhaps most interesting result of using the formalism of currents
directly is the obvious display that this process has \textbf{just a
  $t$-channel pole} (i.e.~no poles in the $s$ or $u$-channel), exactly like
the seemingly simpler $qQ$-scattering process. The pure $t$-channel
structure, and the same colour factors, hold true for \emph{all} the helicity
assignments, where the gluon does not flip helicity during the
scattering. Not only are the helicity-flipping amplitudes systematically
suppressed by the centre-of-mass energy, half of these amplitudes even
still have only a $t$-channel pole. $s$ and $u$ channel poles appear only in
the helicity scatterings where the helicity of the incoming $q$ and $g$ is
identical, \emph{and} the helicity of the gluon is flipped under the
scattering.

In the case of pure gluon scattering, it could seem a little arbitrary to
discuss the $s,t,$ and $u$-channels. However, in the cases of scattering of
two gluons of opposite helicities, like $g^- g^+\to g^-g^+$ it turns out
again that \textbf{the scattering amplitude has just a $t$-channel pole}, and
is again just the contraction of two currents with special colour factors,
which depend only on the acceleration of each gluon during the scattering.

The brief summary presented here of the studies in
Ref.\cite{Andersen:2009nu,Andersen:2009he,Andersen:2010ch,Andersen:2010ih}
illustrates how the $t$-channel exchange is completely well-defined (for
$2\to 2$ processes) not just for $qQ$-scattering, but also for scattering
processes involving gluons. This is displayed in a formalism based on
helicity amplitudes and currents, without resorting to kinematic
approximations or limits. 

\boldmath
\subsection{All Orders Real Corrections}
\label{sec:all-orders-real}
\unboldmath 
The previous section demonstrated that the $t$-channel pole of the full
scattering amplitudes is much more important for the accurate description of
the scattering processes than the zoo of Feynman diagrams would suggest. We
demonstrated how this $t$-channel pole can be described exactly for many
processes by a formalism based on the scattering of specific helicity
currents. For example, the colour summed and averaged amplitudes for both
$qQ\to qQ$ and the dominant ones of $qg\to qg$ scattering (all helicity
non-flipping, like $q^-g^+\to q^-g^+$) are described exactly by a formalism
of pure quark current scatterings, with colour factors depending on the
flavour (quark or gluon) of the scattered partons.

In the current section we will describe briefly the approximations to the
real, radiative corrections of the $2\to 2$ process in \emph{High Energy
  Jets}. The soft and collinear regions are already well understood by the
description in a parton shower. HEJ focuses instead on the hard, radiative corrections. The
aim is to build a framework which is sufficiently accurate for a ``first
guess'' for the impact of the radiative corrections (i.e.~to all order with a
certain logarithmic accuracy), but which then is also sufficiently flexible
to include matching to the full fixed-order result, where this is
accessible. The control of the cross section to leading logarithmic accuracy
in $\log(s/t)$ requires control of the hard scattering matrix element to
leading power in $s/t$, as $s/t\to\infty$. As discussed in the previous
section, and in more depth in
e.g.~Ref.\cite{Andersen:2008gc,Andersen:2009nu,Andersen:2009he}, the control
of the leading power alone is achieved already in the formalism of
Fadin-Kuraev-Lipatov\cite{Fadin:1975cb,Kuraev:1976ge,Kuraev:1977fs}, but this
is insufficient to ensure a good description of the scattering amplitudes in
the energy regime of the LHC. The formalism described here combines the right
limit at $s/t\to\infty$ (or more generally the MRK limit) with complete
gauge-invariance. It is already clear that the description of the
$2\to2$-processes discussed in the previous section is gauge invariant, since
it describes the $t$-channel pole of the full scattering amplitude exactly. In
the current section we will build a gauge-invariant approximation to $2\to
n$-processes. 

\subsubsection{Dominant $n$-jet configurations}
\label{sec:domin-conf}

First, we will discuss briefly which processes dominate the $2\to n$
partonic scattering in the MRK limit. 
For any $2\to n$ scattering process, the final state particles can obviously
be ordered according to rapidity. Apart from exceptional phase space
points (of zero measure), no two particles will have the same rapidity.

At the currently implemented accuracy, the HEJ amplitudes will describe the
leading contribution (in the invariant mass between two neighbouring partons)
to the $n$-jet production process. For a given $n$-jet kinematic
configuration, many of the possible partonic channels will be systematically
suppressed.
These channels will not be summed to all orders, but will be included ``only''
through matching corrections. Consider now the rapidity ordered final state
jets. The leading contributions to the $n$-jet configurations are those
where the flavour of the most forward jet equals that of the incoming parton
of positive light-cone momentum, and the flavour of the most backward jet is
identical to that of the incoming parton of negative light-cone momentum. The
leading contribution to jet production between the jets extremal in
rapidity is given by pure gluon emissions. Such processes can proceed through
a gluon exchange between all rapidity-ordered particles. Changing the
flavours of two jets, such that a single gluon propagator between the two jets is
replaced by a $t$-channel quark propagator, automatically leads to a
suppression of $1/s_{ij}$ for $s_{ij}\to\infty$, where $s_{ij}$ is the
invariant mass between the two jets. We choose to call the leading configurations
\emph{FKL}-configurations, since they are the same as those which are considered
in the amplitudes by Fadin-Kuraev-Lipatov~\cite{Fadin:1975cb,Kuraev:1976ge,Kuraev:1977fs,Balitsky:1979ap,DelDuca:1993pp,DelDuca:1995zy}.

\subsubsection{Amplitudes and Effective Vertices}
\label{sec:ampl-effect-vert}
The all-order approximations of the $n$-parton FKL-configurations are
constructed similarly to the $2\to2$ scattering
amplitudes considered in Section~\ref{sec:dominance-t-channel}, as effective
vertices connected by $t$-channel propagators. In the case of the partons of
largest or smallest rapidity, these are directly the effective currents
discussed in the previous section. The emission of additional gluons is
performed by gauge-invariant\footnote{by which we of course mean fully gauge
  invariant, not just up to sub-asymptotic terms as it is often meant in the
  BFKL literature.}, effective vertices. These were derived in
Ref.\cite{Andersen:2009nu}, and take into account the leading contribution
from emissions off both the $t$-channel exchange and the two incoming and the
most forward/backward outgoing partons.  The effective vertex for the
emission of a gluon of momentum $p_g=q_1-q_2$, $V^\mu(q_i,q_{i+1})$, is given
by\cite{Andersen:2009nu} 
\begin{align}
  \label{eq:GenEmissionV}
  \begin{split}
  V^\rho(q_i,q_{i+1})=&-(q_i+q_{i+1})^\rho \\
  &+ \frac{p_A^\rho}{2} \left( \frac{q_i^2}{p_{i+1}\cdot p_A} +
  \frac{p_{i+1}\cdot p_B}{p_A\cdot p_B} + \frac{p_{i+1}\cdot p_n}{p_A\cdot p_n}\right) +
p_A \leftrightarrow p_1 \\ 
  &- \frac{p_B^\rho}{2} \left( \frac{q_{i+1}^2}{p_{i+1} \cdot p_B} + \frac{p_{i+1}\cdot
      p_A}{p_B\cdot p_A} + \frac{p_{i+1}\cdot p_1}{p_B\cdot p_1} \right) - p_B
  \leftrightarrow p_n.
  \end{split}
\end{align}
This form of the effective vertex is fully gauge invariant; the Ward
Identity, $p_g\cdot V=0$ can easily be checked. This allows for a meaningful
approximation to the scattering amplitude to be constructed.

\begin{figure}
  \hspace{6cm}
  \epsfig{file=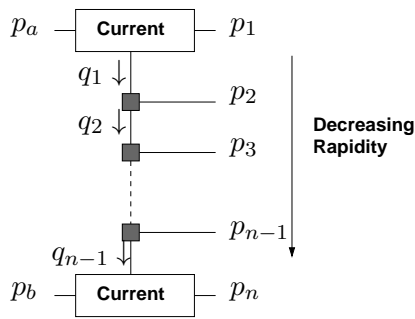,width=0.3\textwidth}
  \hspace{-5.5cm}
  \put(0.25,38.){$p_a$}
  \put(0.25,3.){$p_b$}
  \put(29,38.){$p_1$}
  \put(29,29.){$p_2$}
  \put(29,22.){$p_3$}
  \put(29,11.){$p_{n-1}$}
  \put(29,3.){$p_n$}
  \put(9.,31.5){$q_1\downarrow$}
  \put(9.,25.){$q_2\downarrow$}
  \put(5.5,7.5){$q_{n-1}\downarrow$}
  \caption{The analytic structure of a scattering amplitude in \emph{High
      Energy Jets}.}
  \label{fig:HEJstructure}
\end{figure}

Another approximation of HEJ is then the systematic omission of interference
effects between identical particles, since such effects are suppressed by the
invariant mass between the particles. Essentially, each emission is treated
as a distinguishable particle, just like in a parton shower. The resulting
tree-level approximation for a $2\to n$ scattering is illustrated in
Fig.~\ref{fig:HEJstructure}. Virtual corrections modify the $t$-channel
propagators and are discussed together with regularisation in the next
section.
The tree-level HEJ-approximation for the square of the amplitude describing a
$qQ$-scattering process with $n$ jets in the final state is then given
by\cite{Andersen:2009nu}
\begin{align}
  \label{eq:multijetVs}
  \begin{split}
    \left|\overline{\mathcal{M}}^t_{qQ\to qg\ldots gQ}\right|^2\ =\ &\frac 1 {4\
      (\Nc^2-1)}\ \left\|S_{qQ\to qQ}\right\|^2\\
    &\cdot\ \left(g^2\ \Cf\ \frac 1 {t_1}\right) \cdot\ \left(g^2\ \Cf\ \frac 1
      {t_{n-1}}\right)\\
    & \cdot \prod_{i=1}^{n-2} \left( \frac{-g^2 C_A}{t_it_{i+1}}\
      V^\mu(q_i,q_{i+1})V_\mu(q_i,q_{i+1}) \right),
  \end{split}
\end{align}
where $\left\|S_{qQ\to qQ}\right\|^2$ indicates the square of pure current-current
scattering of Sec.~\ref{sec:dominance-t-channel}.  In the case of scattering of gluons, the terms in this sum are
weighted with helicity-dependent colour factors\cite{Andersen:2009he}, one of
which is listed in Eq.~\eqref{eq:sqcol}. All the building blocks for
constructing the \emph{High Energy Jets}-scattering amplitudes are listed in
Appendix~\ref{cha:build-blocks-regul}.

\subsection{All Orders Virtual Corrections}
\label{sec:all-orders-virtual}
The virtual corrections are approximated with the \emph{Lipatov ansatz} for
the $t$-channel gluon propagators (see Ref.\cite{Andersen:2009nu} for more
details). The \emph{Lipatov Ansatz} states that order by order, the
leading logarithmically virtual corrections to the full $n$-parton scattering
amplitude in the MRK limit can be obtained by the following replacement in 
the scattering amplitudes:
\begin{align}
  \label{eq:LipatovAnsatz} \frac 1 {t_i}\ \to\ \frac 1 {t_i}\ \exp\left[\hat
\alpha (q_i)(y_{i-1}-y_i)\right]
\end{align} with
\begin{align} 
  \hat{\alpha}(q_i)&=-\gs^2\ \Ca\
  \frac{\Gamma(1-\varepsilon)}{(4\pi)^{2+\varepsilon}}\frac 2
  \varepsilon\left({\bf q}^2/\mu^2\right)^\varepsilon.\label{eq:ahatdimreg}
\end{align} 
This ansatz for the exponentiation of the virtual corrections in the
appropriate limit of the $n$-parton scattering amplitude has been proved to
even the sub-leading
level\cite{Bogdan:2006af,Fadin:2006bj,Fadin:2003xs,Fadin:2005pt}. In
Sec.~\ref{sec:running-coupling} we will discuss parts of the next-to-leading
logarithmic corrections, which can be included as corrections of the type
$\beta_0 \log(Q^2/\mu^2)$.

\subsection{Generation and Regularisation of the Cross Section}
\label{sec:regul-cross-sect}
We will now discuss the construction of the all-order, regularised dijet cross
section. The necessary details for constructing a generator were already given in Ref.\cite{Andersen:2008gc,Andersen:2009nu}, but
the discussion here is more detailed. We begin by defining the matrix element
squared built from the $t$-channel factorised picture (eq.~\eqref{eq:multijetVs})
combined with the virtual corrections discussed in the previous section:
\begin{align}
  \label{eq:Mtvepsilon}
  \begin{split}
    \overline{\left|\mathcal{M}^{t,v}_{\varepsilon\ f_1 f_2 \to f_1 g\cdot g f_2}\right|}^2 = \ &\frac 1 {4\
       (\Nc^2-1)}\ \left\|S_{f_1 f_2\to f_1 f_2}\right\|^2\\
     &\cdot\ \left(g^2\ K_{f_1}\ \frac 1 {t_1}\right) \cdot\ \left(g^2\ K_{f_2}\ \frac 1
       {t_{n-1}}\right)\\
     & \cdot \prod_{i=1}^{n-2} \left( \frac{-g^2 C_A}{t_it_{i+1}}\
       V^\mu(q_i,q_{i+1})V_\mu(q_i,q_{i+1}) \right)\\
     & \cdot \prod_{j=1}^{n-1} \exp\left[2 \hat
         \alpha (q_j)(y_{j-1}-y_j)\right],
  \end{split}
\end{align}
where $f_1, f_2$ indicate the flavour (quarks or gluon), $S_{f_1 f_2\to f_1 f_2}$ is the
sum of contracted currents, and $K_{f_1}$ is $\Cf$ if $f_1=q$ and $\Ca$ if $f_1=g$.  These
pieces are all given explicitly in Appendix~\ref{cha:build-blocks-regul}.

The dijet inclusive cross section is simply constructed as the explicit phase space
integral over the explicit sum of real, radiative corrections, including the leading,
all-order virtual corrections. We illustrate the procedure with $qQ$-scattering, but the
generalisation to incoming gluons is straightforward using the gluon currents and factors
detailed in Appendix~\ref{cha:build-blocks-regul}.
\begin{align}
  \begin{split}
    \label{eq:dijetunreg}
    \sigma_{qQ\to2j}=&\sum_{n=2}^\infty\ \prod_{i=1}^n\left(\int
    \frac{\mathrm{d}^2\mathbf{p}_{i\perp}\ \mathrm{d}y_i}{2\ (2\pi)^3}\right)\
  \frac{\overline{|\mathcal{M}^{t,v}_{\varepsilon\ f_1 f_2\to f_1 g\cdot gf_2}|}^2}{\hat
    s^2} \ x_a f_{A,q}(x_a, Q_a)\ x_2 f_{B,Q}(x_b, Q_b) 
     \\
    &\times(2\pi)^4\ \delta^2\!\!\left(\sum_{k=1}^n \mathbf{p}_{k\perp}\right )\
    \mathcal{O}_{2j}(\{p_i\}). 
  \end{split}
\end{align}
Here, $(p_{i\perp},y_i)$ denote the transverse momentum and rapidity of the
$i$'th final state parton. The parton momenta fractions are given by
\begin{align}
  x_a=\sum_{i}\frac{|p_{i\perp}|}{\sqrt s} \exp(-y_i), \qquad
  x_b=\sum_{i}\frac{|p_{i\perp}|}{\sqrt s} \exp(y_i)\label{eq:bjorkenxs},
\end{align}
with $\sqrt{s}$ the total hadronic centre-of-mass energy. In
Eq.~\eqref{eq:dijetunreg}, $\sqrt{\hat s}$ denotes the total partonic
centre-of-mass energy, $\hat s= x_a x_b s$, and $f_{A,q}(x_a, Q_a),
f_{B,Q}(x_B, Q_B)$ denote the relevant parton density functions for parton
$A,B$ respectively at the resolution scales $Q_A, Q_B$. We will discuss the
choices of scales further in Sec.~\ref{sec:running-coupling}. The function
$\mathcal{O}_{2j}(\{p_i\})$ takes as arguments all the final state partons,
and returns 1 if there are at least two jets, according to the chosen
jet-definition. It is otherwise zero. In the current study, we choose to apply the anti-kt 
algorithm as implemented
in \textsc{FastJet}\cite{Cacciari:2008gp}, with a $R$-parameter of 0.6;
however, 
obviously any jet-definition can be applied on the partonic ensemble. 

The integration over transverse momentum runs from 0 to infinity. We choose
to generate only the rapidity ordered phase space
(i.e.~$y_{i-1}<y_i<y_{i+1}$) using the approach of
Ref.\cite{Andersen:2006sp}, since the \emph{HEJ}-amplitudes
$\overline{|\mathcal{M}^{t,v}_\varepsilon|}^2$ take as argument the rapidity ordered set
$\{p_i\}$. The phase space integration of standard fixed-order amplitudes can
be done in a similar way (and indeed is done in the matching-procedure of
Sec.~\ref{sec:matching}), where then an additional Monte Carlo sampling is
performed over the identification between the particle leg and the rapidity
ordered set of momenta. The phase space generation method of
Ref.\cite{Andersen:2006sp} is very efficient for processes dominated by
$t$-channel poles.

The matrix elements $\overline{|\mathcal{M}^{t,v}_\varepsilon|}^2$ are
divergent for any $p_{i\perp}\to 0$. We will first discuss how for all but the
extremal partons, this divergence cancels with the pole in $\varepsilon$ from
the virtual corrections implemented according to the Lipatov Ansatz for the
resummed $t$-channel propagators (we will then return to the case of the
extremal partons below).

Consider the limit where the transverse momentum of the $i$th
emitted gluon is vanishing. In this limit,
\begin{align}
  \label{eq:kisoftlimit}
  \overline{\left|{\mathcal{M}}^{t,v}_{\varepsilon\ p_a\ p_b\to p_1\ \cdots\ p_{i-1}\ p_i\ p_{i+1}\ \cdots\ 
    p_n}\right|}^2\ \stackrel{{{\bf p}_{i}}^2\to 0}{\longrightarrow}\
\left(\frac{4\ \gs^2\ \Ca}{{{\bf p}_{i}}^2}\right)
  \overline{\left|{\mathcal{M}}^{t,v}_{\varepsilon\ p_a\ p_b\ \to\ p_1\ \cdots\ p_{i-1}\
      p_{i+1}\ \cdots\ p_n}\right|}^2,
\end{align}
where the matrix element on the RHS has $n-1$ final state particles, and
${\bf p}_i^2$ is the sum of the squares of the transverse components of $p_i$
in the Euclidean metric. By integrating over the soft region ${\bf
  p}_i^2<\lambda^2$ of phase space in $D=4+2\varepsilon$ dimensions we find
\begin{align}
  \begin{split}
    \label{eq:realdiv1}
    \int_0^\lambda& \frac{\mathrm{d}^{2+2\varepsilon}{\bf p}\
      \mathrm{d}y_i}{(2\pi)^{2+2\varepsilon}\ 4\pi} \left(\frac{4 \gs^2
        \Ca}{{\bf p}^2}\right)\mu^{-2\varepsilon}\\
    &=\frac{4\gs^2\Ca}{(2\pi)^{2+2\varepsilon}4\pi}\Delta y_{{i-1},{i+1}}
    \frac{\pi^{1+\varepsilon}}{\Gamma(1+\varepsilon)} \frac 1 \varepsilon (\lambda^2/\mu^2)^\varepsilon.
  \end{split}
\end{align}

The square of the matrix element on the left hand side of
Eq.~(\ref{eq:kisoftlimit}) contains the exponential $\exp(2\hat\alpha(q_i)\Delta
y_{{i-1},{i+1}})$. By expanding the exponential to first order in $\gs^2$ and in
$\varepsilon$, the resulting pole in $\varepsilon$ does indeed cancel that of
Eq.~(\ref{eq:realdiv1}), and the combined effect of one soft real emission
and the first term in the expansion of the Reggeised propagator is a factor
\begin{align}
  \label{eq:virtualexponent}
  \Delta y_{{i-1},{i+1}}\frac{\as\Nc}{\pi} \ln\left(\frac{\lambda^2}{{\bf q}^2}\right)
\end{align}
multiplying the $(n-1)$-particle matrix element. It is clear that the nested
rapidity integrals of additional soft radiation in the $t$-channel factorised
multi-parton amplitudes will build up the exponential needed to cancel the
poles from the virtual corrections to all orders in \as. The divergence
arising from a given real emission is therefore cancelled by that arising
from the virtual corrections in the Reggeised $t$-channel propagator of the
matrix element without the relevant real emission. Therefore, if indeed
Eq.~\eqref{eq:kisoftlimit} had been an equality for $p_i^2<\lambda^2$, then
the regularised HEJ matrix element squared would be:
\begin{align}
  \label{eq:almostMHEJ}
  \begin{split}
    \overline{\left|\mathcal{M}^\mathrm{reg}(\{ p_i\})\right|}^2 = \ &\frac 1 {4\
       (\Nc^2-1)}\ \left\|S_{f_1 f_2\to f_1 f_2}\right\|^2\\
     &\cdot\ \left(g^2\ K_{f_1}\ \frac 1 {t_1}\right) \cdot\ \left(g^2\ K_{f_2}\ \frac 1
       {t_{n-1}}\right)\\
     & \cdot \prod_{i=1}^{n-2} \left( \frac{-g^2 C_A}{t_it_{i+1}}\
       V^\mu(q_i,q_{i+1})V_\mu(q_i,q_{i+1}) \right)\\
     & \cdot \prod_{j=1}^{n-1} \exp\left[\omega^0(q_j,\lambda)(y_{j-1}-y_j)\right], \\
     \omega^0(q_j,\lambda)=\ &-\frac{\alpha_s N_C}{\pi} \log\frac{{\bf q}_j^2}{\lambda^2},
  \end{split}
\end{align}
which should only be evaluated for $\mathbf{p}_i^2>\lambda^2$, and a simple
phase-space slicing would then have been sufficient to organise the
cancellation of divergences. However, while Eq.~\eqref{eq:kisoftlimit} does
describe the divergence in the soft limit, it is not an exact identity. We
can account for the finite difference by including an integration over
\begin{align}
  \label{eq:subtraction}
  \frac {-1}{t_i t_{i+1}}V^\mu(q_i,q_{i+1})V_\mu(q_i,q_{i+1}) - \frac{4}{\mathbf{p}_i^2}
\end{align}
for $\mathbf{p}_i^2<\lambda^2$. Numerically, it turns out to be sufficient to
account for the difference and include this integral for values of
$|\mathbf{p}_i|$ above roughly 0.2~GeV. The regulated matrix elements for
\emph{HEJ} are then given by
\begin{align}
  \label{eq:MHEJ}
  \begin{split}
    \overline{\left|\mathcal{M}_{\rm HEJ}^\mathrm{reg}(\{ p_i\})\right|}^2 = \ &\frac 1 {4\
       (\Nc^2-1)}\ \left\|S_{f_1 f_2\to f_1 f_2}\right\|^2\\
     &\cdot\ \left(g^2\ K_{f_1}\ \frac 1 {t_1}\right) \cdot\ \left(g^2\ K_{f_2}\ \frac 1
       {t_{n-1}}\right)\\
     & \cdot \prod_{i=1}^{n-2} \left( {g^2 C_A}\
       \left(\frac {-1}{t_it_{i+1}} V^\mu(q_i,q_{i+1})V_\mu(q_i,q_{i+1}) -
         \frac{4}{\mathbf{p}_i^2}\ \theta\left(\mathbf{p}_i^2<\lambda^ 2\right)\right)\right)\\
     & \cdot \prod_{j=1}^{n-1} \exp\left[\omega^0(q_j,\lambda)(y_{j-1}-y_j)\right], \\
     \omega^0(q_j,\lambda)=\ &-\frac{\alpha_s N_C}{\pi} \log\frac{{\bf q}_j^2}{\lambda^2}.
  \end{split}
\end{align}
Since the $t$-channel factorised matrix elements are very fast to
evaluate and the regularisation procedure does not add any complexity
(because of the simple IR structure of the $t$-channel factorised matrix
elements), the radiative corrections to all orders can be constructed as an
explicit phase space integral over each number of gluons emitted\footnote{The
lower limit on the transverse momentum in the phase space integrals is
understood to be small, but non-zero, so Eq.~\eqref{eq:subtraction} can still
be evaluated numerically.}: 
\begin{align}
  \begin{split}
    \label{eq:dijetreg}
    \sigma_{qQ\to2j}=&\sum_{n=2}^\infty\
    \prod_{i=1}^n\left(\int_{p_{i\perp}=0}^{p_{i\perp}=\infty}
      \frac{\mathrm{d}^2\mathbf{p}_{i\perp}}{(2\pi)^3}\ 
      \int \frac{\mathrm{d} y_i}{2}
    \right)\
    \frac{\overline{|\mathcal{M}_{\mathrm{HEJ}}^\mathrm{reg}(\{ p_i\})|}^2}{\hat s^2} \\
    &\times\ \ x_a f_{A,q}(x_a, Q_a)\ x_2 f_{B,Q}(x_b, Q_b)\ (2\pi)^4\ \delta^2\!\!\left(\sum_{k=1}^n
      \mathbf{p}_{k\perp}\right )\ \mathcal{O}_{2j}(\{p_i\}).
  \end{split}
\end{align}
The cancellation of the poles in $\varepsilon$ ensures that the logarithmic
dependence on $\lambda$ generated by the effective lower limit on the transverse
momentum integrals cancels with the logarithmic $\lambda$-dependence of the
virtual + unresolved-real correction, which generates the exponential factors
of Eq.~\eqref{eq:virtualexponent}. This is similar to the explicit
construction of the solution to the BFKL evolution, where the very weak
dependence of the solution on $\lambda$ at leading logarithmic accuracy was
studied in Ref.~\cite{Schmidt:1996fg,Orr:1997im}, and in
Ref.~\cite{Andersen:2003wy} at next-to-leading logarithmic accuracy. In
Appendix~\ref{sec:vari-regul-param} we investigate the stability under
variations in $\lambda$ of a few of the cross-sections and distributions
discussed throughout this paper. We find that
the residual $\lambda$-dependence is very weak --- see
Sec.~\ref{sec:vari-regul-param} for further details. We then generally choose
to use $\lambda=0.5$~GeV. Note that these findings are in good agreement with
the conclusions from the studies of the $\lambda$-dependence of the explicit
solutions to the BFKL
equations\cite{Schmidt:1996fg,Orr:1997im,Andersen:2003wy,Andersen:2004uj},
where the convergence of the phase space integration could be checked
explicitly against an analytic solution.

\begin{figure}[htbp]
  \centering
  \epsfig{width=.8\textwidth,file=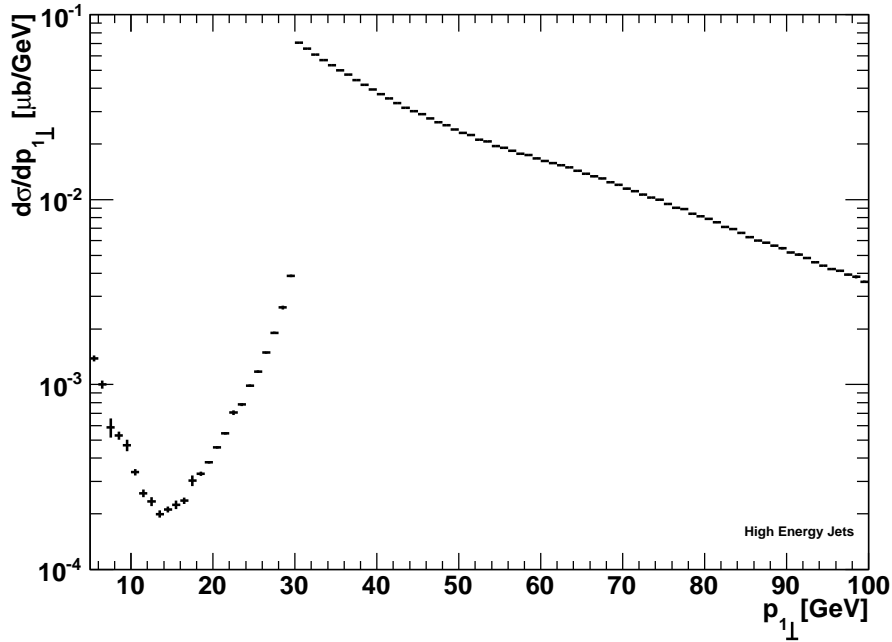}
  \caption{The dependence of the dijet cross section on the transverse
    momentum of the parton of lowest or highest rapidity. The step at 30GeV
    is caused by the requirement of minimum 30GeV transverse momentum by the
    hard jets.}
  \label{fig:ptadep}
\end{figure}

The only remaining unregulated divergences of
$\overline{|\mathcal{M}_\mathrm{HEJ}^\mathrm{reg}|}^2$ are related to the
region of zero transverse momenta of the partons extremal in
rapidity\footnote{Actually, with the emission vertex of
  Eq.~\eqref{eq:GenEmissionV} there is also a collinear divergence for
  emissions close to the extremal partons from parts symmetrising
  $p_A\leftrightarrow p_1$ and $p_B\leftrightarrow p_n$. We avoid this
  divergence by not
averaging over the two contributions for emissions which are clustered into
the same jets as the extremal partons.}. A similar situation 
was discussed in
Ref.\cite{Andersen:2008gc}, where simply a cut on the transverse momentum of
the extremal partons was introduced, and the dependence of the cross section
on this cut studied. We have refined the treatment for the current study. If
there is no hard jet associated with the extremal partons, they could be
viewed as not participating in the proper hard scattering of the event. In
the parton shower picture, such emissions would be counted as (in this case)
initial state radiation, and the divergence regulated by the Sudakov form
factors. The treatment of these are beyond the scope of the current paper,
and we will simply require that the extremal partons are associated with
(i.e.~a member of) a hard jet. With this requirement, the dependence on a
lower cut-off of the momentum allowed for the extremal partons is weak. This
is illustrated in Fig.~\ref{fig:ptadep} for a dijet-sample at the 7~TeV LHC,
requiring just\footnote{Note that such a simple cut is problematic for NLO
  studies, because the truncation of the perturbative series introduces a
  large logarithmic dependence on any difference in the value of the cut applied on the
  two jets\cite{Frixione:1997ks}.} two anti-kt-jets with
absolute rapidities less than 4.5, and with transverse momenta above 30~GeV.
We see that the contribution from transverse momenta much smaller than the
jet scale is small. The requirement that the extremal partons be associated
with a hard jet has to a large extent regulated the divergence for vanishing
transverse momentum of the extremal partons (compare with Fig.~16 of
Ref.\cite{Andersen:2008gc}). In the results discussed in
Sec.~\ref{sec:dijet-event-results}, we will choose a lower limit on the
transverse momentum of the extremal partons, which is 5~GeV smaller than the
minimum transverse momentum required on hard jets.  Removing the very small
contribution from smaller scales simply improves the phase space
integration. Furthermore, the unregulated divergence at zero transverse
momentum has to be explicitly removed.

The construction of an explicit integration over emissions to all orders
relies on an efficient phase-space
generator~\cite{Andersen:2008gc,Andersen:2008ue}, which should sample final
states with the number of particles varying by more than an order of
magnitude. The problem is significantly different to that of a so-called
general purpose Monte Carlo (Pythia\cite{Sjostrand:2007gs},
Herwig\cite{Bahr:2008pv}, SHERPA\cite{Gleisberg:2008ta}), since in these
approaches, the approximation to the virtual corrections is \emph{defined}
such that the emission of particles is \emph{unitary}, i.e.~does not change
the total cross section, which allows for a simple probabilistic
interpretation. In \emph{HEJ}, an approximation to the virtual
corrections is calculated, and introduces a \emph{suppression} of the
regularised matrix element for any final state with a finite number of
partons, as the rapidity length of the event is increased. This is countered
by the (positive) contribution from the emission of additional gluons, and
introduces a correlation between the number of final state partons and the
typical rapidity length of an event. It is absolutely crucial to incorporate this probabilistic correlation 
in the phase space generator in order to obtain satisfactory numerical
stability in a finite amount of time. Such a phase space integrator can be
efficiently implemented by following the ideas of
Ref.\cite{Andersen:2006sp}. The fully exclusive formulation in a
flexible Monte Carlo facilitates the study of any observable.


\section{Matching}
\label{sec:matching}
The previous sections have set up the all-order approximations to jet production of
\emph{High Energy Jets}, and discussed the implementation as a flexible Monte Carlo,
integrating explicitly over $n$-particle phase space. The resummation procedure generates
only certain partonic phase space configurations (\emph{FKL}-configurations, see Section
\ref{sec:domin-conf}). The dijet production process is calculated within this
approximation as (for notational brevity, we have omitted the label indicating the use of the
regularised amplitudes)
\begin{align}
  \begin{split}
    \label{eq:resumdijet}
    \sigma_{2j}^\mathrm{resum}=&\sum_{f_1, f_2}\ \sum_{n=2}^\infty\
    \prod_{i=1}^n\left(\int_{p_{i\perp}=0}^{p_{i\perp}=\infty}
      \frac{\mathrm{d}^2\mathbf{p}_{i\perp}}{(2\pi)^3}\ 
      \int \frac{\mathrm{d} y_i}{2}
    \right)\
    \frac{\overline{|\mathcal{M}_{\mathrm{HEJ}}^{f_1 f_2\to f_1 g\cdots gf_2}(\{ p_i\})|}^2}{\hat s^2} \\
    &\times\ \ x_a f_{A,f_1}(x_a, Q_a)\ x_2 f_{B,f_2}(x_b, Q_b)\ (2\pi)^4\ \delta^2\!\!\left(\sum_{k=1}^n
      \mathbf{p}_{k\perp}\right )\ \mathcal{O}_{2j}(\{p_i\}),
  \end{split}
\end{align}
where the first sum is over the flavours $f_1, f_2$ of incoming partons. The
distribution of any observable can be obtained by simply binning the cross
section in Eq.~\eqref{eq:resumdijet} in the appropriate variable formed from
the explicit momenta. Obviously, multi-jet rates can also be calculated by
multiplying by further multi-jet observables $\mathcal{O}_{3j},
\mathcal{O}_{4j},\ldots$ in Eq.~\eqref{eq:resumdijet}.

In section~\ref{sec:match-fkl-conf} we will discuss how the amplitudes for
the FKL-states included in Eq.~\eqref{eq:resumdijet} can be corrected to full
tree-level accuracy, limited only by the availability of full tree-level
matrix elements. In Section \ref{sec:matching-non-fkl} we will discuss the
inclusion of all remaining partonic configurations (in practice for up to 4 jets).

\subsection{Matching for FKL Configurations}
\label{sec:match-fkl-conf}
Firstly, we want to match the description of the FKL $n$-jet configurations to
the full tree-level matrix elements and thus improve upon the approximations
inherent to the resummation. This can be straightforwardly done
because of the flexibility inherent in Eq.~\eqref{eq:resumdijet}. Let
$\mathcal{O}_{nj}^e(\{p_i\})$ denote the measurement function for \emph{exclusive}
$n$-jet production acting on the partonic phase space. This function will
return one if the chosen jet-algorithm finds exactly $n$ hard jets in the
$m$-partonic phase space point, and returns zero otherwise. 
Furthermore, it will give access to the
momenta of the $n$ jets, $\{p_{\mathcal{J}_l}(\{p_i\})\}$. We note that
$\mathcal{O}_{2j}(\{p_i\})=\sum_{n=2}^\infty\mathcal{O}^e_{nj}(\{p_i\})$. In
principle, we would then want to simply multiply each exclusive jet measure
function with
\begin{align}
  \label{eq:reweightfactorjets}
  \frac{\overline{|\mathcal{M}^{f_1f_2\to f_1g\cdots gf_2}\left(\left\{p_{\mathcal{J}_l}(\{p_i\})\right\}\right)|}^2}{\overline{|\mathcal{M}^{t,f_1f_2\to f_1g\cdots gf_2}(\{p_{\mathcal{J}_l}(\{p_i\})\})|}^2},
\end{align}
where the numerator is simply the (spin and colour summed and averaged)
square of the full $n$-jet tree-level matrix element, and the denominator is the
\emph{HEJ}-approximation to this tree-level. This would ensure tree-level
accuracy of the $n$-jet rates, while simultaneously weighing the $n$-jet
samples with the virtual corrections from \emph{HEJ}.

However, a few modifications to this na\"ive approach are necessary. Firstly, the jet
momenta may not be of zero invariant mass. Secondly, the transverse momenta of the jets
generally will not sum to zero, since some of the partons generated in the event may not
be included in the hard jets. We therefore have to construct a new set of $n$ jet-momenta
to be used in the matching. We start by making each jet momentum equal to the sum of the
parton momenta of each jet (each jet contains mostly just one hard parton after the
\emph{HEJ}-resummation). We then redistribute the transverse momenta of any partons not
belonging to a jet among the hard jets, and remove these softer partons from the list of
particles (and momenta) used in the matching.  We choose to distribute the momenta in
proportion to the transverse momenta of the resolved jets. If the sum of the momenta of
the non-jet partons is $q$ and the scalar sum of the transverse momenta of the jets is
$P_\perp$, the new set of hard momenta $p^{\rm new}_{\mathcal{J}_l}$ is given by
\begin{align}
  \label{eq:ptreassign}
  p^\mathrm{new}_{\mathcal{J}_l} = p_{\mathcal{J}_l} + q *
  \frac{|p_{\mathcal{J}_{l\perp}}|}{P_\perp}.
\end{align}
The energy component of each jet is then finally reset to put it on-shell, and the
momenta of the incoming partons are defined by energy/momentum conservation.

\begin{figure}[btp]
  \centering
  \epsfig{file=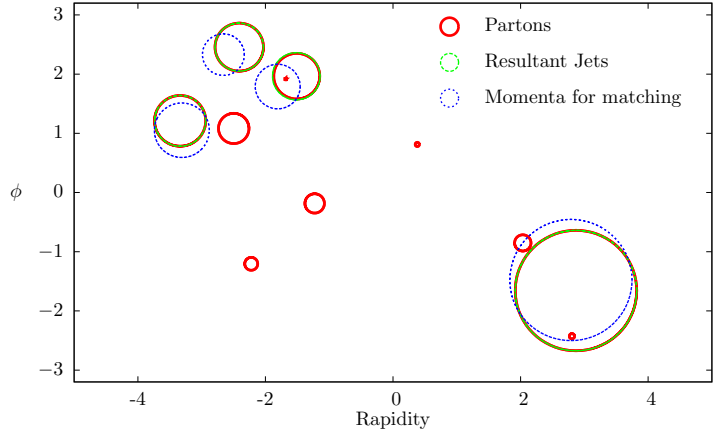,width=0.6\textwidth}  
  \caption{This plot shows for an example event the momenta of the partons
    (red), the resultant jets from FastJet (green) and the reshuffled momenta
    described in the text (blue).  The radii of the circles are proportional
    to the transverse momentum of the particle or jet described.}
  \label{fig:pteffect}
\end{figure}

This reshuffling of momenta is illustrated for a sample event in
figure~\ref{fig:pteffect}, which has eleven partons in the final
state, in a momentum configuration leading to four hard jets with transverse
momentum above 30~GeV, found with the anti-kt jet algorithm, as implemented in
\textsc{FastJet}\cite{Cacciari:2008gp}.
The red circles show the positions in rapidity-phi space of the partons; 
the radii of the circles are proportional to the transverse energy of
each parton and jet (and do not, therefore,
represent the area of each jet).
The green circles indicate the jets of the original event. As expected, they
coincide with the hardest quarks/gluons. The blue circles indicate the
reshuffled momenta used in the matching. Note, this procedure does not change the
kinematics of the actual event; only the reweighing of the event to full
tree-level accuracy is performed with matrix elements evaluated for the
slightly modified momenta. If the threshold on the transverse momenta of jets
was set very low, and the jets were finely resolved (small $R$-parameter),
then no reshuffling of momenta would be necessary. However, the full matrix
elements can only be evaluated for states of relatively low multiplicity
(with MadGraph\cite{Alwall:2007st}, we limit ourselves to matching of up to
four jets). So with a low jet matching scale, the available fixed order
matrix elements for matching would cover only a small part of the total cross
section. A similar issue occurs for the
CKKW-L\cite{Catani:2001cc,Lonnblad:1992tz} or MLM\cite{Mangano:2001xp} style
matching of parton shower algorithms.

We then reweigh each event generated with the following multiplicative
matching factor, evaluated with the on-shell hard momenta as found by the
described procedure:
\begin{align}
  \label{eq:matchfact}
  w_{n-\mathrm{jet}}\equiv\frac{\overline{\left|\mathcal{M}^{f_1f_2\to f_1g\cdots
          gf_2}\left(\left\{p^\mathrm{new}_{\mathcal{J}_l}(\{p_i\})\right\}\right)\right|}^2}{\overline{\left|\mathcal{M}^{t,f_1f_2\to
          f_1g\cdots
          gf_2}\left(\left\{p^\mathrm{new}_{\mathcal{J}_l}(\{p_i\})\right\}\right)\right|}^2}.
\end{align}
In this notation, we have suppressed the flavour and momentum-dependence of
$w_n$, but it is obviously calculated on an event-by-event basis. The
FKL-matched cross section is then found as
\begin{align}
  \begin{split}
    \label{eq:resumdijetFKLmatched}
    \sigma_{2j}^\mathrm{resum, match}=&\sum_{f_1, f_2}\ \sum_{n=2}^\infty\
    \prod_{i=1}^n\left(\int_{p_{i\perp}=\lambda}^{p_{i\perp}=\infty}
      \frac{\mathrm{d}^2\mathbf{p}_{i\perp}}{(2\pi)^3}\ 
      \int \frac{\mathrm{d} y_i}{2}
    \right)\
    \frac{\overline{|\mathcal{M}_{\mathrm{HEJ}}^{f_1 f_2\to f_1 g\cdots gf_2}(\{ p_i\})|}^2}{\hat s^2} \\
    &\times\ \sum_m \mathcal{O}_{mj}^e(\{p_i\})\ w_{m-\mathrm{jet}}\\
    &\times\ \ x_a f_{A,f_1}(x_a, Q_a)\ x_2 f_{B,f_2}(x_b, Q_b)\ (2\pi)^4\ \delta^2\!\!\left(\sum_{i=1}^n
      \mathbf{p}_{i\perp}\right )\ \mathcal{O}_{2j}(\{p_i\}).
  \end{split}
\end{align}

The impact of this matching procedure can be seen in Fig.~\ref{fig:FLKmatchydif}, which
displays the differential dijet cross section wrt.~the rapidity difference $\Delta y_{fb}$
between the most forward/backward hard jet, within the following set of cuts:
\begin{align}
  \label{eq:dijetcuts}
  p_{j_\perp}>60\mathrm{GeV}\quad |y_j| < 4.5\quad \mathrm{anti-kt}, R=0.6.
\end{align}
The matching scale is set equal to the general jet scale of 60~GeV. The red (dot-dashed)
curve is the result of the pure resummation; the blue (dashed) curve is obtained after
matching of the states arising in the resummation up to four hard jets.  The correction is
small throughout, being slightly more significant at low rapidity spans.

\subsection{Matching for Non-FKL Configurations}
\label{sec:matching-non-fkl}
\begin{figure}[btp]
  \centering
  \epsfig{file=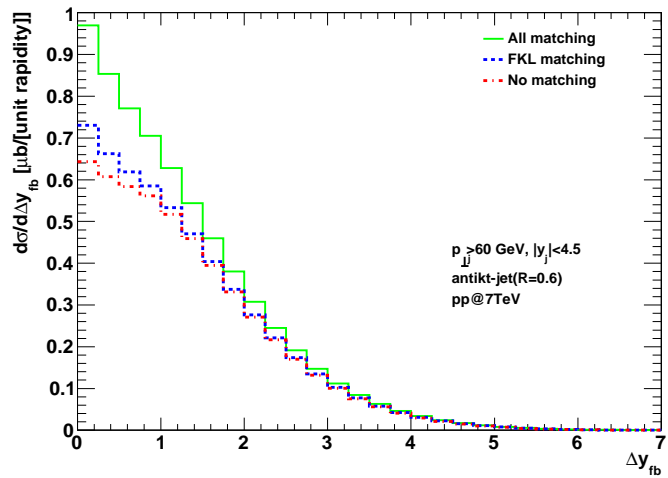,width=0.6\textwidth}
  \caption{This plot shows the impact of matching in jet production, as a function of
    rapidity span.  The initial HEJ approximation is shown (red, dot-dashed) together with
    matching to FKL configurations (blue, dashed) and matching to both FKL
    and non-FKL configurations (green, solid).}
  \label{fig:FLKmatchydif}
\end{figure}

The processes and partonic configurations which do not arise in the
resummation are included straightforwardly by adding these to the dijet rate
found by the (matched) Eq.~\eqref{eq:resumdijet}. For example, we can add the
remaining contribution to the exclusive dijet rate as
\begin{align}
    \label{eq:2jnonfkl}
    \sigma^{\mathrm{non-FKL}}_{2j} =& \sum_{f_1, f_2}\sum_{f_{f1}, f_{f2}}\ \prod_{i=1}^2\left(\int_{p_{i\perp}=p_{\perp\mathrm{min}}}^{p_{i\perp}=\infty}
      \frac{\mathrm{d}^2\mathbf{p}_{i\perp}}{(2\pi)^3}\ \int \frac{\mathrm{d}
        y_i}{2} \right)\
    \frac{\overline{|\mathcal{M}^{f_1 f_2\to f_{f1} f_{f2}}(\{
        p_i\})|}^2}{\hat s^2}\ \\
    &\times\  \Theta(\{f_i\},\{p_i\})\ x_a f_{A,f_1}(x_a, Q_a)\ x_2 f_{B,f_2}(x_b, Q_b)\ (2\pi)^4\
    \delta^2\!\!\left(\sum_{i=1}^n \mathbf{p}_{i\perp}\right )\
    \mathcal{O}_{2j}(\{p_i\}),\nonumber
\end{align}
where $p_{\perp\mathrm{min}}$ is the minimum transverse momentum required for
hard jets. The function $\Theta(\{f_i\},\{p_i\})$ returns one if the parton
and momentum configuration is of non-FKL status. If only rapidity ordered
sets of momenta $p_i$ are generated, then one needs to also sum (or Monte
Carlo sample) over all possible assignments between momenta and the particles
in the process.  The generalisation to the three and four jet states is
straightforward, and the final result for the dijet rate is
\begin{align}
  \label{eq:dijettotal}
  \sigma_{2j}=\sigma_{2j}^\mathrm{resum, match} + \sum_n \sigma^{\mathrm{non-FKL}}_{nj}.
\end{align}
Each component is implemented by explicit Monte Carlo sampling over phase
space and an evaluation of matrix elements. Therefore, any observable can be
constructed and studied, also after matching has been included in the
formalism. 

The impact of the non-FKL states is indicated on Fig.~\ref{fig:FLKmatchydif}, where the
green (solid) line is obtained from the sum of all terms in
Eq.~\eqref{eq:dijettotal}.  This correction is again more significant for small
rapidity spans, as we expect.


\section{Logarithmic Corrections to the Scale Choice}
\label{sec:running-coupling}
The discussions so far have made no assumptions on the scale choice made for
the evaluation of $\alpha_s$ or the pdfs. In this section we will compare the
results arising for a fixed scale choice (of e.g.~the minimum jet transverse
momentum), and a scale choice made event by event equal to the maximum jet
transverse momentum of the event. Finally, we will include pieces of the
next-to-leading logarithmic corrections to the BFKL kernel, which will
stabilise the dependence on the scale choice. This will then form the basis
of the standard scale choice for the results presented in
Section~\ref{sec:dijet-event-results}.

The connection between the formalism of \emph{High Energy Jets} and that of
BFKL\cite{Balitsky:1979ap,Fadin:1975cb,Kuraev:1976ge,Kuraev:1977fs,Fadin:1998py}
is that in the limit of large invariant mass between all partons (conditions
relaxed for neighbouring pairs of particles at NLL\cite{Fadin:1996nw}), then
the amplitudes underlying the BFKL formalism coincide with those of
\emph{HEJ} (and with those of full QCD). 
The NLL corrections to the BFKL kernel have two origins: the one-loop
corrections to one-gluon emission, and the contribution from two-gluon and
quark--anti-quark-emission in quasi-multi-Regge-kinematics (i.e.~not
necessarily a large invariant mass between the pair of particles). The net
result of the corrections is a sum of an expression with the same functional
form as the LL kernel, multiplied by a running coupling logarithm, and a term
of a more complicated kinematic
structure\cite{Andersen:2003an,Andersen:2003wy}. The relevant
discussion of the regularisation of the NLL corrections to the BFKL kernel
was presented in
Ref.\cite{Andersen:2003an,Andersen:2003wy,Andersen:2004uj}. We repeat it
here, with a notation tailored to the present application.

The NLL BFKL kernel is expressed in terms of a transverse momentum, which is
the transverse momentum of the emitted gluon, or the sum of transverse
momenta of the emitted pair of gluons or quark--anti-quark pair. All other
kinematic dependence is integrated over before arriving at the BFKL
kernel. In $D=4+2\varepsilon$ the BFKL amplitudes obey the following relation
at NLL accuracy (compare with Eq.~\eqref{eq:kisoftlimit})
\begin{align}
  \label{eq:BFKLkisoftlimit}
  \overline{\left|\mathcal{M}^\mathrm{BFKL}_{p_a\ p_b\to p_1\ \cdots\ p_{i-1}\ p_i\ p_{i+1}\ \cdots\ 
    p_n}\right|}^2\ =
\mathcal{K}^r (\mathbf{p}_i)\
  \overline{\left|{\mathcal{M}}^\mathrm{BFKL}_{p_a\ p_b\ \to\ p_1\ \cdots\ p_{i-1}\
      p_{i+1}\ \cdots\ p_n}\right|}^2,
\end{align}
with $\mathcal{K}^r (\mathbf{p}_i)=K^r_\varepsilon (\mathbf{p}_i)+ K^r
(\mathbf{p}_i)$, where $K^r
(\mathbf{p}_i)$ is irrelevant for the current discussion, and
\begin{align}
  \begin{split}
    \label{eq:BFKLrealkernel}
    K^r_\varepsilon (\mathbf{p}_i) =& \frac{4\ g_\mu^2\mu^{-2\varepsilon}\
      \Ca}{\mathbf{p}_i^2} \Bigg[ 1 + \frac {g_\mu^2 \mu^{-2\varepsilon}\
        \Ca\ 
        \Gamma(1-\varepsilon)
      }{(4\pi)^{2+\varepsilon}} \Bigg(
        \beta_0/\Nc \ \frac 1 \varepsilon\ \left\{ 1-
          \left(\frac{\mathbf{p}_i^2}{\mu^2}\right)^\varepsilon
          \left(1-\varepsilon^2\ \frac{\pi^2}6\right) \right\}\\
     &+   \left( \frac{\mathbf{p}_i^2}{\mu^2} \right)^\varepsilon \left( 
       \frac 4 3 - \frac{\pi^2} 3+ \frac 5 3 \frac{\beta_0}\Nc\ +\
       \varepsilon\left(
         14 \zeta(3) - \frac {32}9 -\frac{28}9 \frac{\beta_0}\Nc
       \right)
     \right)\Bigg)
    \Bigg],
  \end{split}
\end{align}
with $\beta_0=\frac{11}3\Nc - \frac 2 3 \nf$. The NLL-corrections to the
trajectory give
\begin{align}
  \begin{split}
    \label{eq:nlltrajectory}
    \hat \alpha(q^2)=&-\bar g_\mu^2\ \frac2 \varepsilon\
    \left(q^2/\mu^2\right)^\varepsilon \Bigg( 1 + \frac{\bar
      g_\mu^2}\varepsilon \Bigg[\left( \beta_0/\Nc\right) \left(1 -
      \frac{\pi^2}6\varepsilon^2\right)\\
    &-\left(\frac{q^2}{\mu^2}\right)^\varepsilon \bigg(\frac{11}6 +
      \left(\frac{\pi^2}6 - \frac{67}{18} \right)\varepsilon +
      \left(\frac{202}{27} - \frac{11 \pi^2}{18} -
        \zeta(3)\right)\varepsilon^2\\
      &-\frac{\nf}{3 \Nc} \left(
        1-\frac 5 3 \varepsilon + \left(
          \frac{28}9 - \frac{\pi^2}3
        \right)\varepsilon^2
      \right) \bigg)\Bigg]\Bigg).
  \end{split}
\end{align}
By applying the same regularisation procedure as discussed in
Sec.~\ref{sec:regul-cross-sect} we find that for both the real emission
(evaluated above $\mathbf{p}_i^2>\lambda^2$) and for the trajectory the term
found at LL accuracy is multiplied by a running coupling logarithm. For
the real emission this is:
\begin{align}
  \label{eq:NLLrealemissionreg}
  \left(\frac{4\ g_\mu^2\ \Ca}{\mathbf{p}_i^2}\right) \left ( 1 - \frac{g_\mu^2}{(4 \pi)}
    \frac{\beta_0}{4\pi}\ln \mathbf{p}_i^2/\mu^2\right).
\end{align}
For the regularised trajectory we find
\begin{align}
  \begin{split}
    \label{eq:NLLtrajectoryrun}
    \omega_0(q,\lambda) =&\ \frac{\alpha_s\
      \Ca}{\pi}\ln\left(\frac{\lambda^2}{q^2}\right)\ \left(
      1+\frac{\alpha_s}2 \frac{\beta_0}{4 \pi}\ln\frac{\mu^4}{q^2\lambda^2}
    \right).
  \end{split}
\end{align}
These results are in complete agreement with what was found in
Ref.\cite{Andersen:2003an,Andersen:2003wy}. The logarithm of the trajectory may seem a
little odd (being dependent on $\lambda$), but it reproduces the NLL BFKL
results when expanded in $\beta_0$. Besides, the study of the pure NLL BFKL correction in
Ref.\cite{Andersen:2003an,Andersen:2003wy,Andersen:2005jr}, show that the
organisation of the cancellation of soft divergence is completely stable for the values
explored for $\lambda$.

We finish off this section with a simple study illustrating the impact of
various scale choices on the average number of hard jets versus the rapidity
difference between the most forward/backward jet within the cuts of
Eq.~\eqref{eq:dijetcuts}. We apply three different choices: 1) a fixed
scale choice of 60GeV, 2) a common scale choice, chosen event by event, of the largest
transverse momentum of any jet, and 3) the latter, including the logarithmic
corrections discussed above. 

\begin{figure}[btp]
  \centering
  \epsfig{file=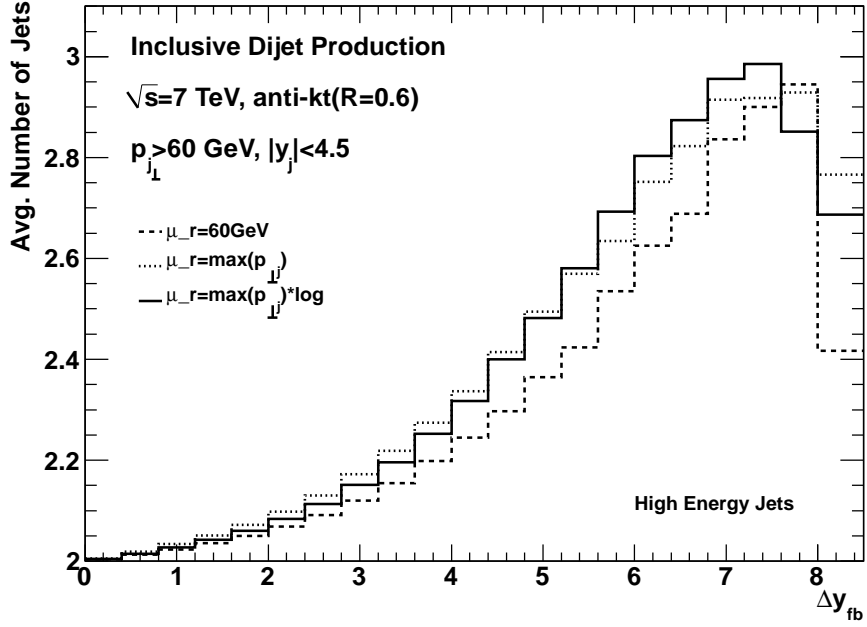,width=0.8\textwidth}
  \caption{The average number of hard jets ($p_{\perp}>60$~GeV) in inclusive
    dijet production as a function of the difference in rapidity $\Delta
    y_{fb}$ between the most forward/backward hard jet.}
  \label{fig:scalechoice}
\end{figure}

This observable is just one of many with a strong
correlation with the number of hard jets -- in order to describe the region of
phase space of large $\Delta y_{fb}$, it is clearly imperative to describe
correctly the emissions of many hard jets. Other such examples are studied in
the next section. The obvious expectation is that for a larger value of
$\alpha_s$ (smaller scale), one would see more hard jets than for a smaller
value of $\alpha_s$ (larger scale). Indeed, this is found in
Fig.~\ref{fig:scalechoice} for rapidity differences less than roughly
5. Furthermore, we see that including the logarithmic corrections outlined above
leads to a prediction in-between that of the fixed, low scale choice of
60GeV, and the choice of the hardest jet scale. This is of course entirely as
expected. For larger rapidities, phase space constraints become increasingly
important, and the scale at which the pdfs are evaluated will influence the
details. Eventually, as $\Delta y_{fb}$ increases further, the average number
of hard jets decreases as the phase space for additional radiation is
reduced when the energy of the forward/backward jets gets close to the total
available hadronic energy.


\section{Results}
\label{sec:dijet-event-results}
In this section we present results for dijet- and trijet-studies within these
cuts (identical to those of Eq.~\eqref{eq:dijetcuts} in
Section~\ref{sec:match-fkl-conf}):
\begin{align}
  \label{eq:dijetcutsagain}
  p_{j_\perp}>60\mathrm{GeV}\quad |y_j| < 4.5\quad \mathrm{anti-kt}, R=0.6.
\end{align}
We choose as the standard scale choice $\mu_r=\max_j(p_{\perp j})$, and
include the running coupling logarithms from NLL accuracy, as discussed in
Section~\ref{sec:running-coupling}. As shown explicitly in
Appendix~\ref{sec:stability-equal-cut}, the all-order framework of \emph{High
  Energy Jets} is free from the instability seen in the NLO-calculation of
dijet production\cite{Frixione:1997ks} when the transverse momentum cut on
the two jets is equal. This problem simply arises from the fact that in a
three-particle system, the cut on the transverse momentum of two particles
automatically changes the phase space explored by the transverse momentum of
the third particle. The infra-red region of the real emission corrections to
the dijet system is explored in the limit where the two transverse momenta of
the hard jets are equal. An off-set $\Delta p_\perp$ in
the cut of the two hardest jets modifies the soft phase space for additional real
emission, and can therefore introduce a logarithmic dependence on $\Delta
p_\perp$. However, this dependence seems specific to cross sections
terminated at NLO, and is washed away in several other all-order frameworks,
e.g.~POWHEG\cite{Nason:2004rx,Alioli:2010xd,Alioli:2010xa} (NLO matched to a
parton shower) and the BFKL generator studied in
Ref.\cite{Andersen:2001kt}. Since the problem is related to a fixed-order
perturbative calculation rather than any observation or our description, we
will proceed with an equal cut on the transverse momentum of all jets.

We will apply the anti-kt jet-clustering algorithm as defined and implemented
in Ref.\cite{Cacciari:2008gp} with $R=0.6$. We define $\Delta y_{fb}$ as the
rapidity difference between the most forward and most backward hard jet. The
average number of jets in the events is an obvious indication of the
importance of the hard, higher order corrections that are resummed in
\emph{High Energy Jets}. In Section~\ref{sec:rapid-distr} we study simple
characteristics of the inclusive sample generated with \emph{HEJ}; we then
move on to discuss distributions in $p_\perp$, $H_T$ (scalar sum of
transverse momenta) and $s_{ij}$ (invariant mass between hardest jets), where the corrections have a
particularly large impact. Other all-order approaches like
e.g.~Cascade\cite{Jung:2000hk,Jung:2010si} calculate higher order corrections
in the $k_t$-factorisation scheme through the evolution of off-shell pdfs
convoluted with a $2\to2$ (off-shell) hard scattering matrix element. It
would be interesting to compare the predictions for these observables also
from such a framework.

\subsection{DiJet Studies}
\label{sec:generic-multi-jet}

\subsubsection{Rapidity and Transverse Momentum Distributions}
\label{sec:rapid-distr}
\begin{figure}[htbp]
  \centering
  \epsfig{width=.49\textwidth,file=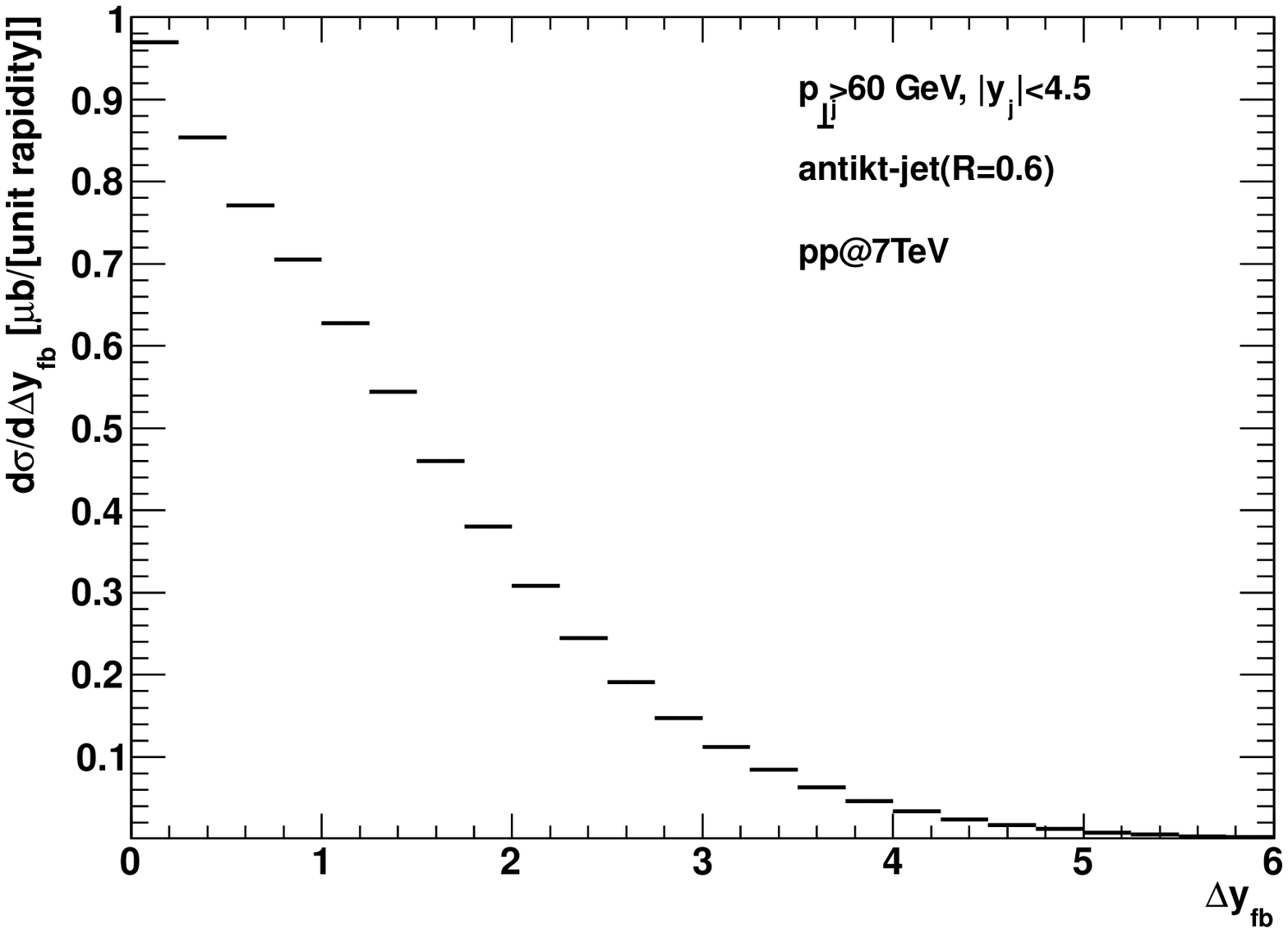}
  \epsfig{width=.49\textwidth,file=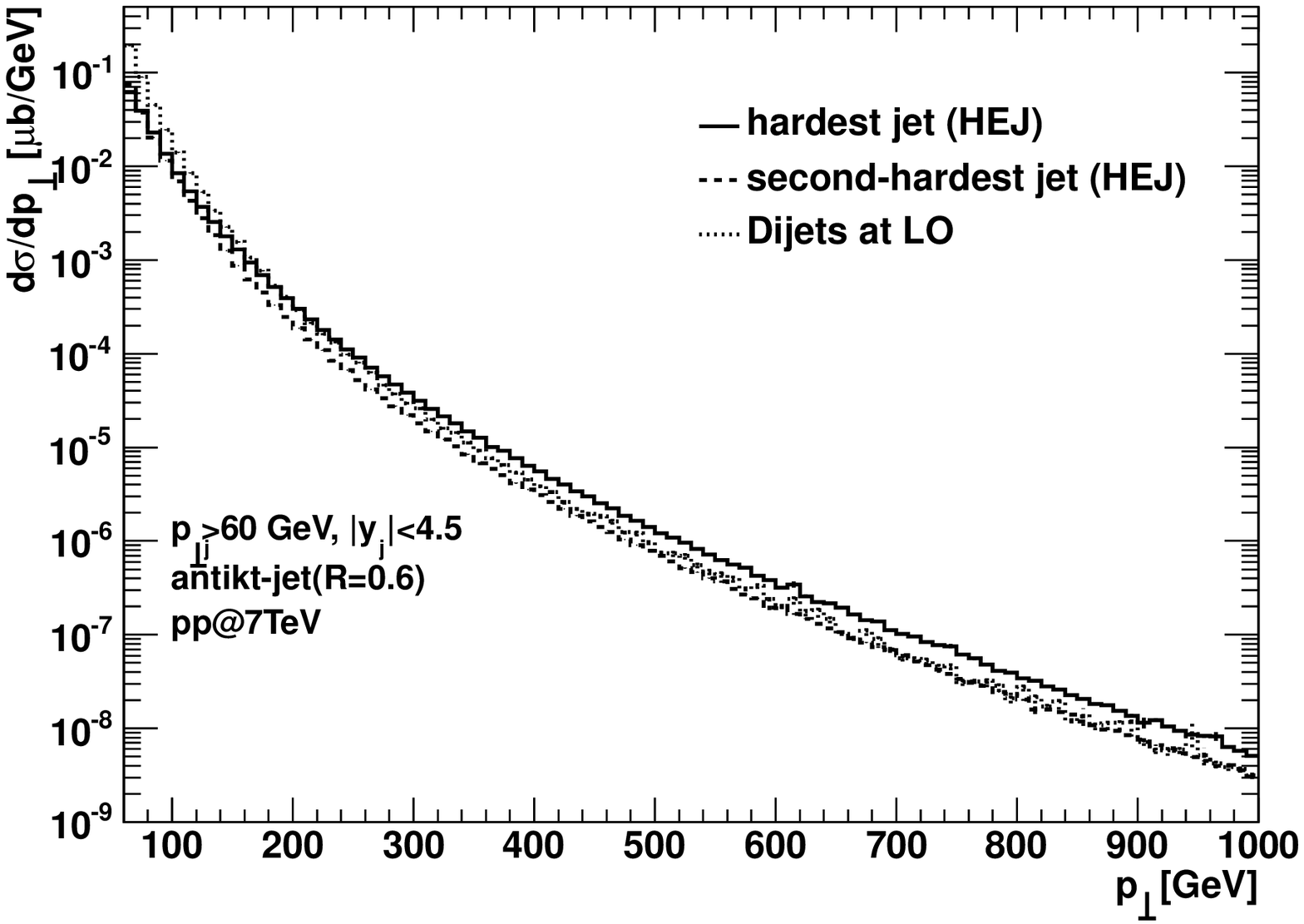}
  \caption{The differential dijet-rate with respect to the rapidity
    difference between the most forward and most backward hard jet (left) and
    the transverse momentum of the hardest and second-hardest jet of the
    event (right).}
  \label{fig:ydif60GeV}
\end{figure}
In Fig.~\ref{fig:ydif60GeV} we have plotted the differential cross section
with respect to both $\Delta y_{fb}$ (left) and the transverse momentum of
the hardest and second-hardest jet in the event (right). The dijet rate is
peaked at zero rapidity difference, and the radiative corrections have
clearly induced a difference in the transverse momentum distribution of the
hardest and second-hardest jet (which is obviously identical at leading
order). The transverse momentum spectrum is compared to that arising in a LO
calculation (using the MSTW2008LO pdf set, and setting the renormalisation
and factorisation scale equal to $p_{\perp j}$). The LO spectrum is
significantly softer then that of the hardest jet arising in \emph{HEJ}.

\subsubsection{ $\Delta y_{fb}$, $H_T$, $s_{j_1 j_2}$ and the Average Number of Jets}
\label{sec:average-number-jets}
In Fig.~\ref{fig:scalechoice} we plotted the average number of hard jets
(transverse momentum larger than 60GeV) according to the anti-kt jet
algorithm with $R=0.6$ in the inclusive dijet sample, as a function of the
rapidity span $\Delta y_{fb}$ between the most forward and most backward hard
jet. As expected, there is a strong correlation between $\Delta y_{fb}$ and
the average number of hard jets. The average number of hard jets rises
monotonously until $\Delta y_{fb}\approx 7$, simply because the partonic
phase space increases. However, as the rapidity span is increased further,
the parton density functions fall off so steeply as $x\to 1$ that the
production of additional hard jets beyond the required dijet system is
effectively vetoed.

\begin{figure}
  \centering
  \epsfig{width=.49\textwidth,file=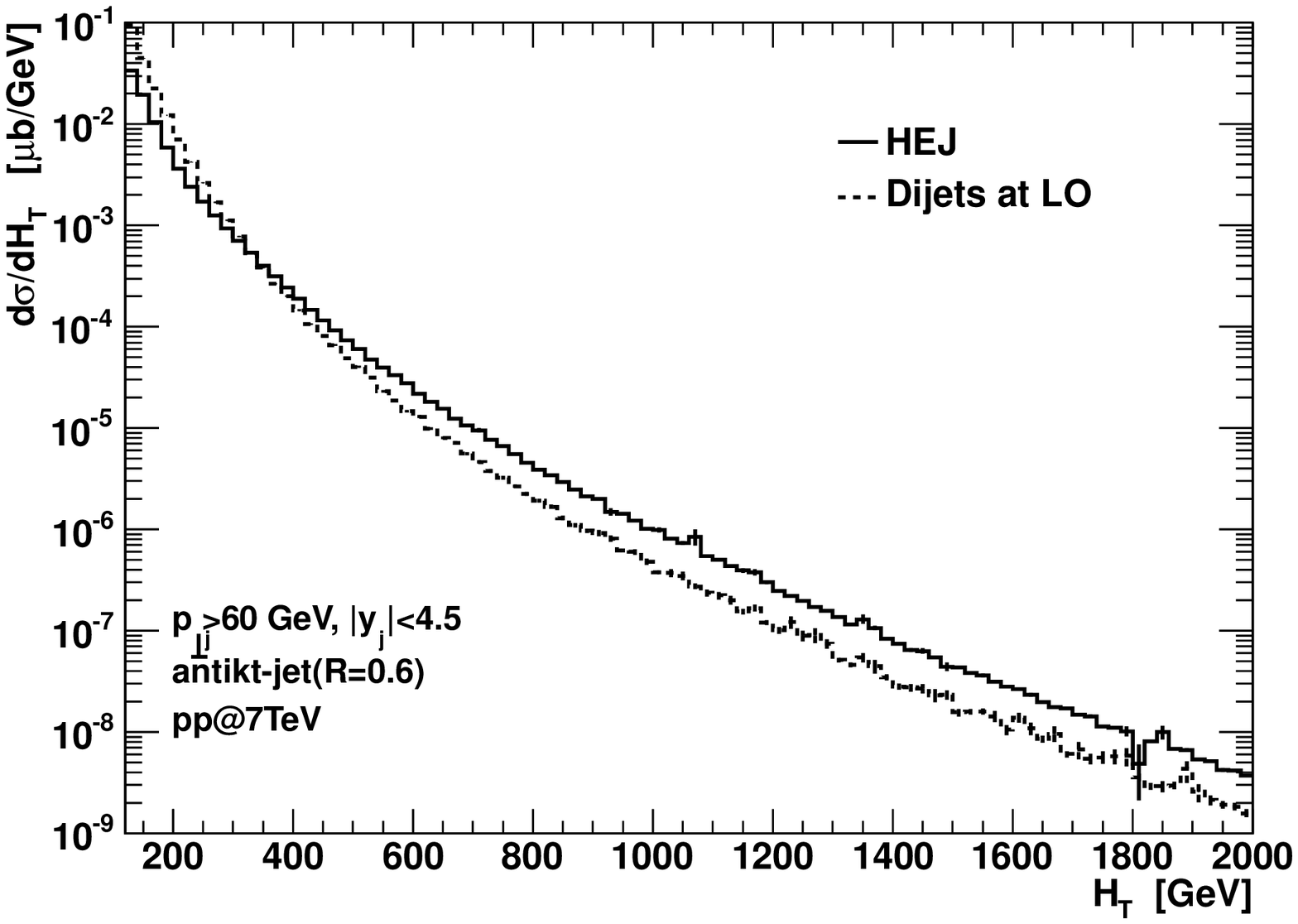}
  \epsfig{width=.49\textwidth,file=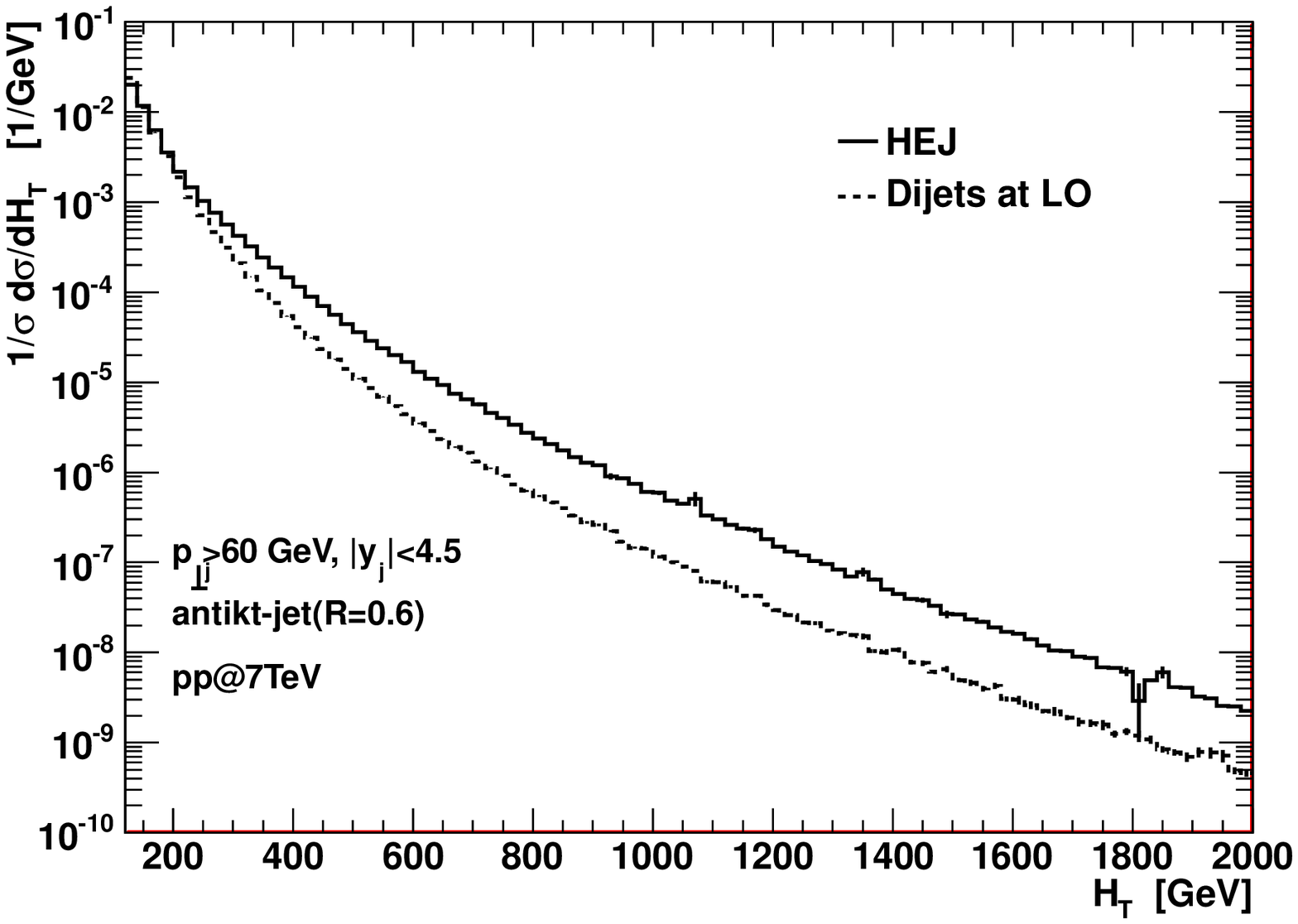}
  \epsfig{width=.49\textwidth,file=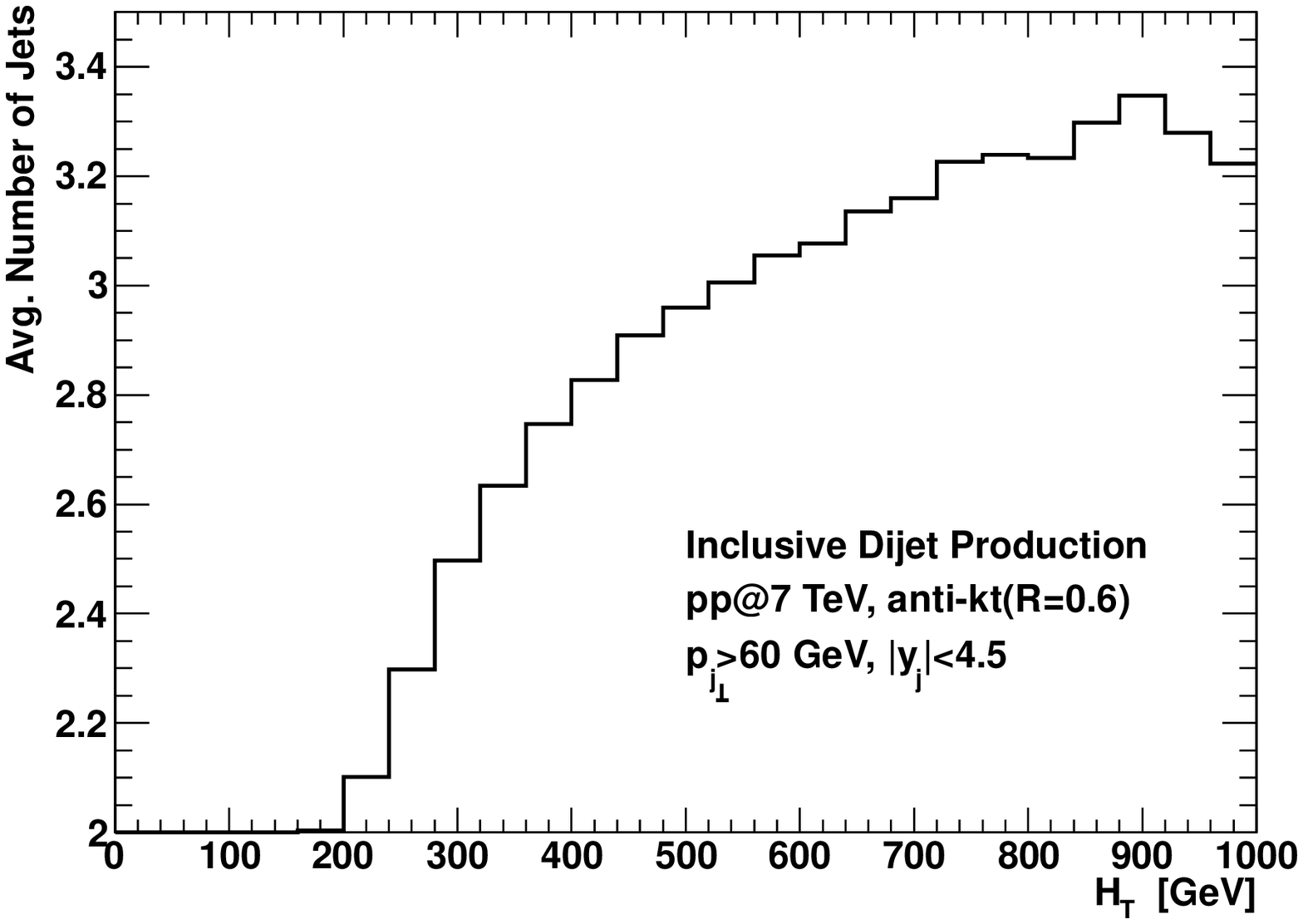}
  \caption{The differential cross section with respect to $H_T$ (top left),
    and the normalised spectrum (top right). The radiative corrections
    implemented in \emph{High Energy Jets} enhance the high-$H_T$-tail
    significantly. The bottom plot is of the average number of hard jets
    (transverse momentum above 60GeV) in the events as a function of
    $H_T$. Hard radiative corrections, as those included in HEJ, are clearly
    important in the description of events with large $H_T$.}
  \label{fig:ht}
\end{figure}
One observable which is often used in the search for signals of new physics
at hadron colliders is the scalar sum of transverse energy (or momentum) in
the hard event. For the jet studies, we define it as
\begin{align}
  \label{eq:ht}
  H_T=\sum_j |p_{\perp j}|,
\end{align}
where the sum runs over the jets found with a given jet-algorithm, with a
transverse momentum bigger than some hard cut-off. In Fig.~\ref{fig:ht} (top
right) we plot the differential cross section wrt.~$H_T$ as obtained both at
leading order QCD, and within \emph{HEJ}. The distribution is clearly more
pronounced at large $H_T$ when the higher order corrections from \emph{HEJ}
are included. This is made very clear on the plots of the normalised
$H_T$-distribution at the top right of Fig.~\ref{fig:ht}. The bottom plot in
Fig.~\ref{fig:ht} is of the average number of jets in the events as a
function of $H_T$. We see that the average number of jets starts at 2, and
very quickly rises above 3 (already at roughly 600GeV). A priori, one might
have expected the large-$H_t$ tail to be dominated by two hard
jets. Fig.~\ref{fig:ht} clearly demonstrates this is not the
case. Furthermore, the very high average number of jets in the large-$H_T$
tail of the dijet distribution suggests that a veto on further hard jets
beyond two would be very efficient in suppressing the QCD contribution to
large-$H_T$ dijet events. The rise in the number of hard jets is a direct
consequence of the $t$-channel colour exchange, and therefore may be
different between the QCD process and any process originating from new physics.

\begin{figure}
  \centering
  \epsfig{width=.49\textwidth,file=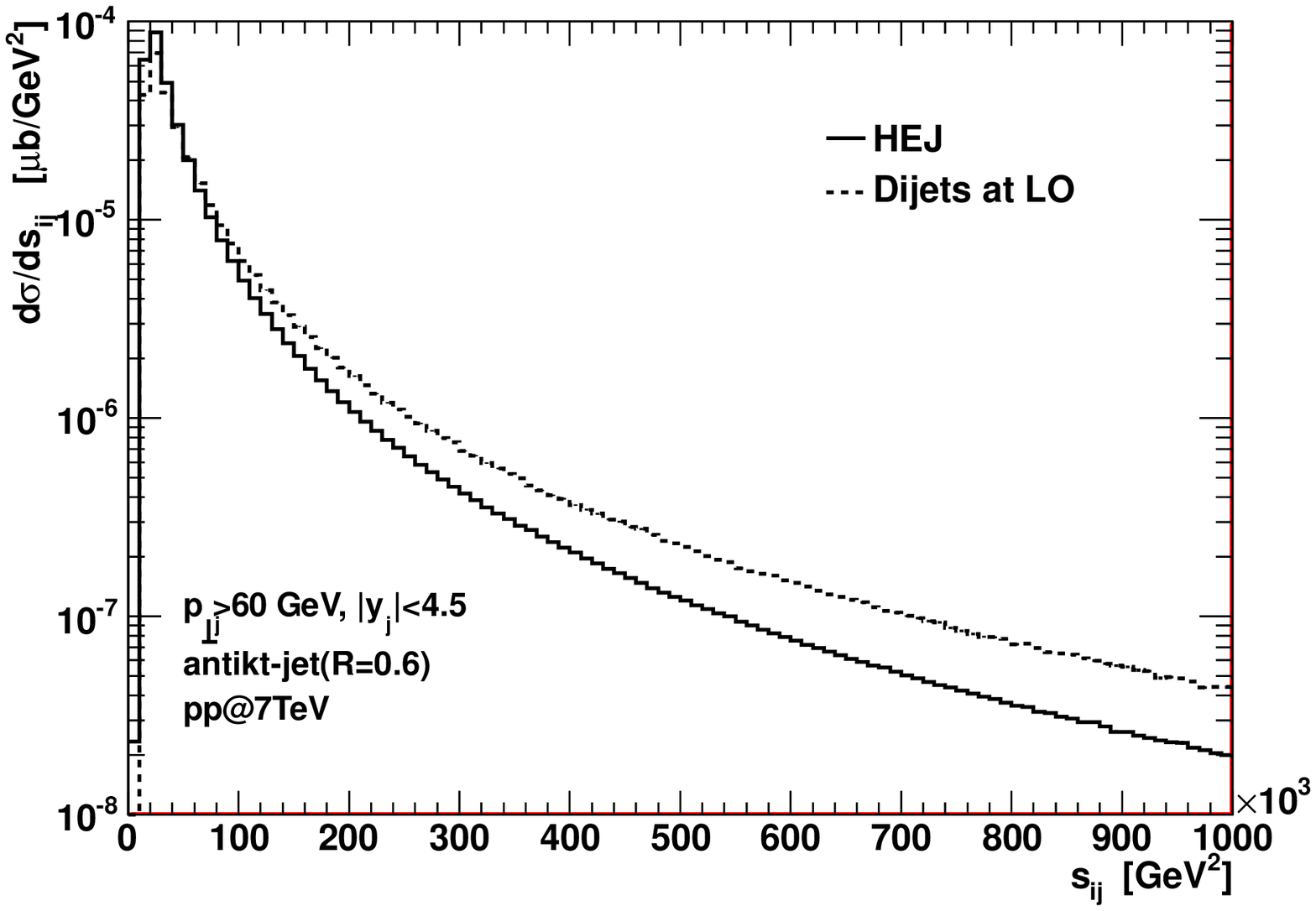}
  \epsfig{width=.49\textwidth,file=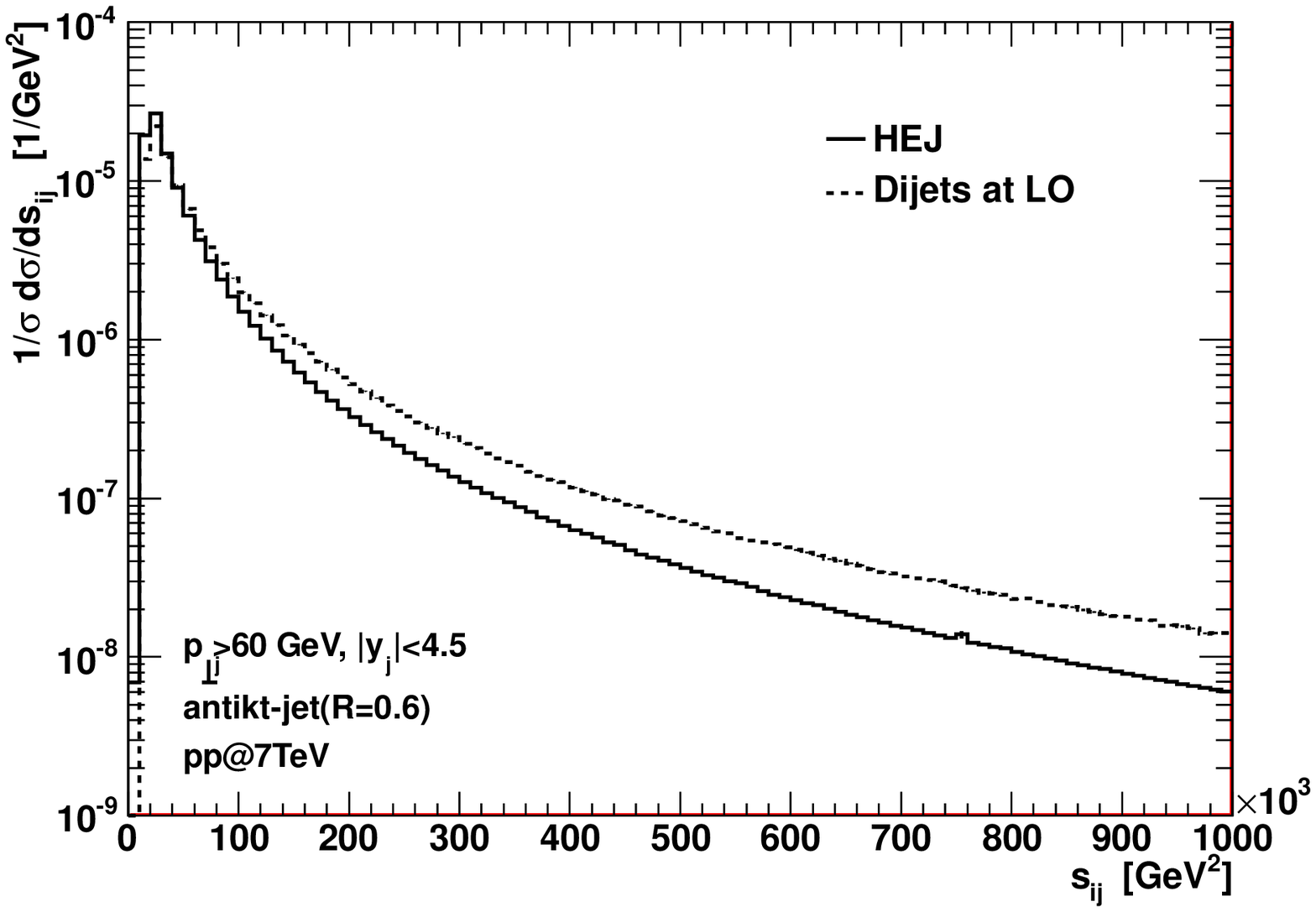}
  \epsfig{width=.49\textwidth,file=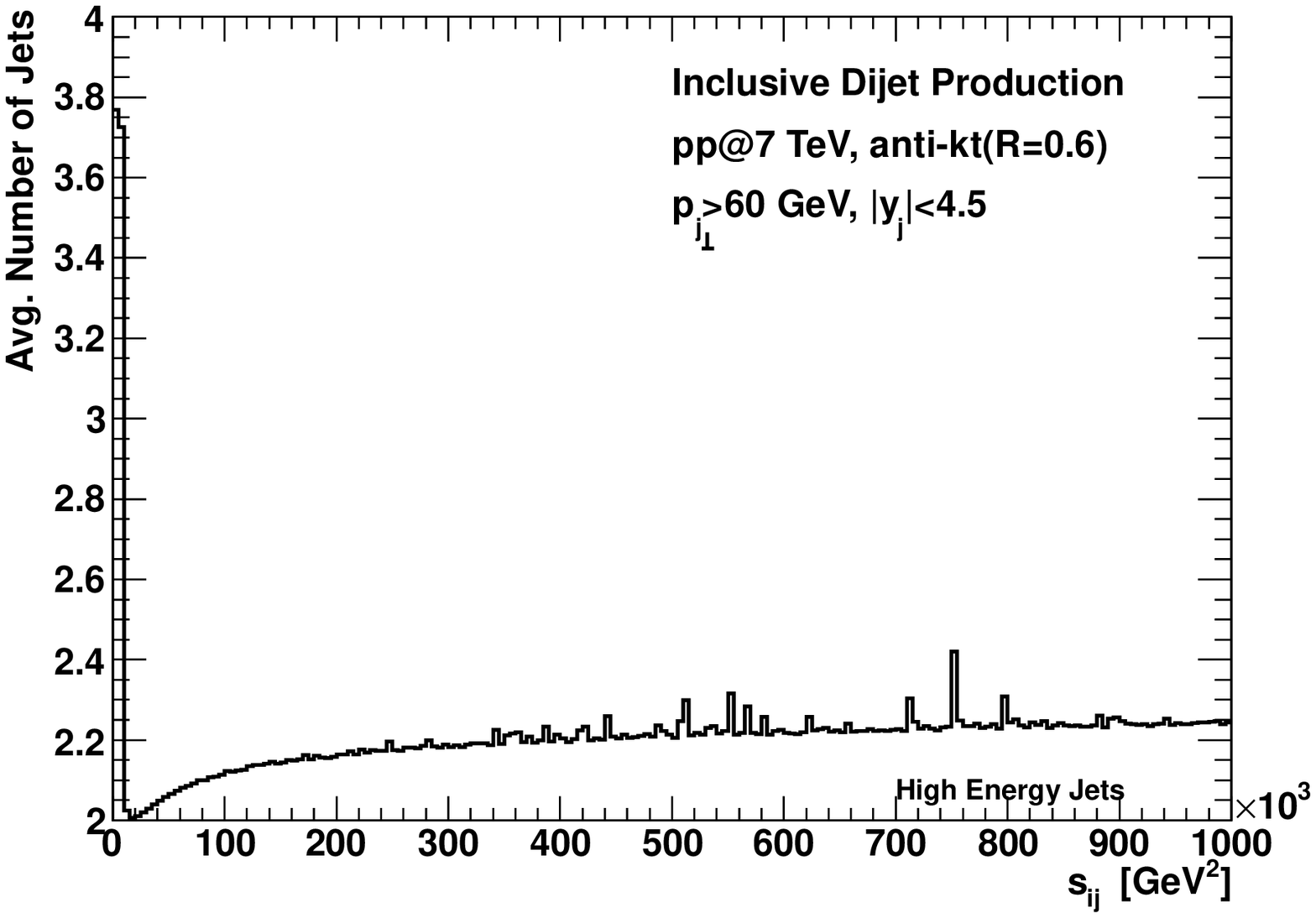}
  \caption{The differential cross section with respect to the square of the
    invariant mass $s_{ij}$ between the two hardest jets (top left), and the
    normalised spectrum (top right). The bottom plot is of the average number
    of hard jets (transverse momentum above 60GeV) in the events as a
    function of $s_{ij}$. Please see text for discussion.}
  \label{fig:sij}
\end{figure}
In Fig.~\ref{fig:sij} we plot the same three quantities for $s_{ij}$, the
square of the invariant mass between the two \emph{hardest} jets of the
event. From the top-right plot of the normalised distribution we see that the
corrections implemented in \emph{HEJ} lead to a relative enhancement at
small $s_{ij}$, and a suppression at large $s_{ij}$ compared to the LO
result. This behaviour can be explained by the fact that the radiative
corrections implemented in \emph{HEJ} will fill the rapidity span between the
two LO jets, as seen in e.g.~Fig.~\ref{fig:scalechoice}. The hardest jets
are likely to be the ones that are radiated centrally in rapidity (a similar
effect was seen in Ref.\cite{Andersen:2008gc}). For any value of the rapidity
span $\Delta y_{fb}$ between the most forward and most backward hard jet, the
higher order corrections implemented in \emph{HEJ} can produce a central jet,
which has a slightly harder transverse momentum spectrum than the extremal
ones, and is therefore more likely to be the one(s) used in the definition of
$s_{ij}$. For all values of $\Delta y_{fb}$, the \emph{HEJ}-corrections will
lead to a smaller value $\Delta y_h$ of the rapidity span between the
two hardest jets in the event. This in turn induces a smaller value of $s_{ij}$
than would be the case in the description of LO exclusive dijets.

The average number of jets versus $s_{ij}$ is shown at the bottom of
Fig.~\ref{fig:sij}. At small values of $s_{ij}$ it is peaked at almost 3.8,
then falls off abruptly to 2 already at $s_{ij}=2\cdot 10^4\mathrm{GeV}^2$,
and then rises to a plateau at 2.25. This behaviour is a sum of two
effects. The events at small $s_{ij}$ are dominated by the cases where two
additional (central) jets have been radiated by the \emph{HEJ}-mechanism;
these two jets are often the hardest (in pt), and therefore define $s_{ij}$
(which can be small since the jets can be aligned in $p_\perp$ (i.e.~they do
not have to be back-to-back in azimuth as dijets at LO) and close in
rapidity). The strong peak at small $s_{ij}$ is therefore an (at least)
$\alpha_s^2$-correction to the tree-level dijets. There is another effect,
giving rise to a distribution increasing with $s_{ij}$, starting at 2 for
$s_{ij}=0$ and then reaching the plateau. This is just the standard
$\mathcal{O}(\alpha_s)$-effect of one hard radiation. The value $2.25$ is not
too far from the value for the average number of jets in the fully inclusive
dijet sample generated with \emph{HEJ}, so the plateau is just an indication
of only a small correlation between $s_{ij}$ and the average number of
jets. However, for the $\mathcal{O}(\alpha_s)$-correction of additional jets
to arise and the average number of jets to rise from the LO value of two, a
rapidity difference between the two jets is required
(c.f.~Fig.~\ref{fig:scalechoice}), and this naturally leads to an increase
in $\hat s_{ij}$. This why the average number of jets is close to 2 only for
small $\hat s_{ij}$. 

\subsection{Trijet studies}
\label{sec:trijet-studies}
\begin{figure}
  \centering
  \epsfig{width=.49\textwidth,file=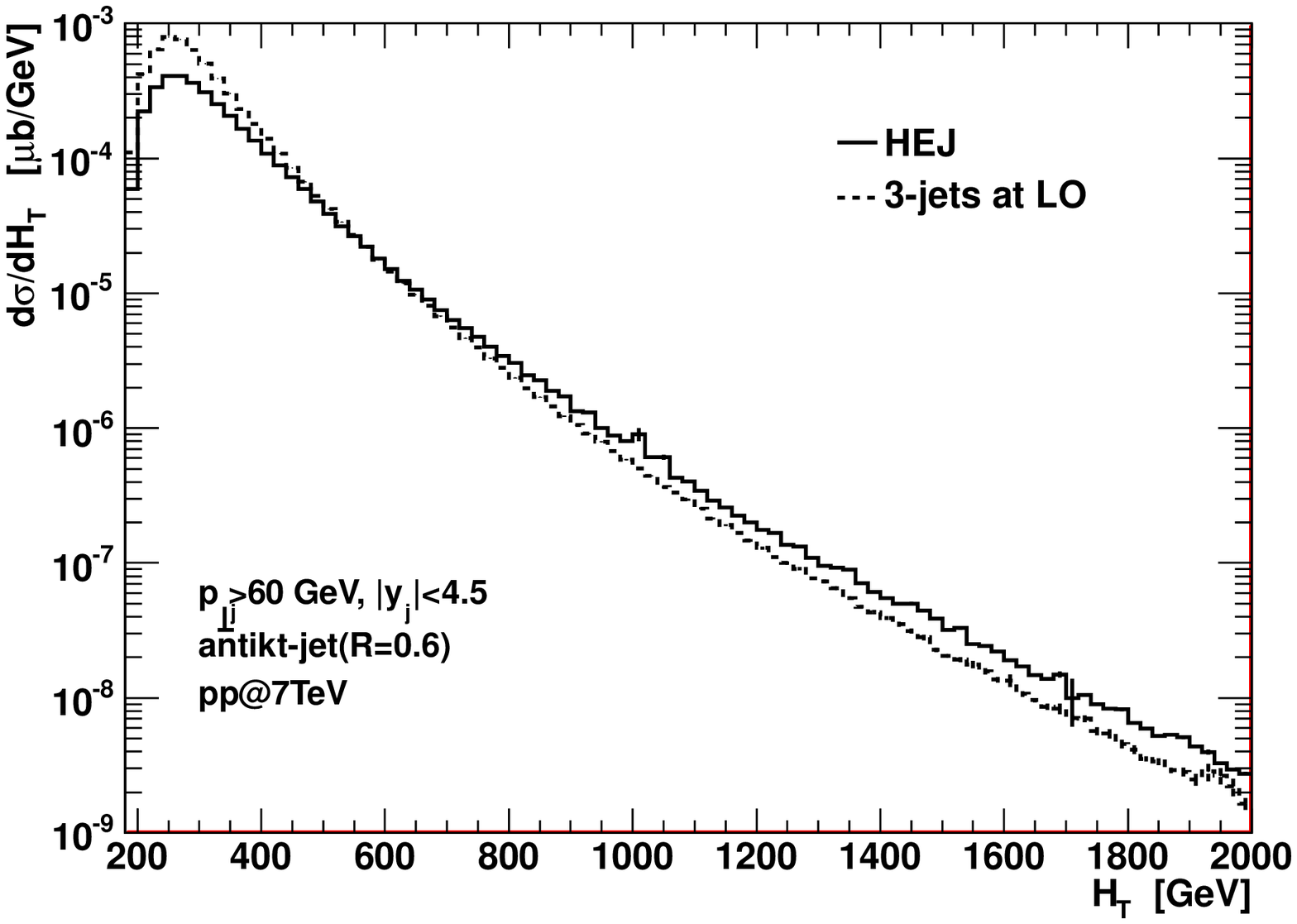}
  \epsfig{width=.49\textwidth,file=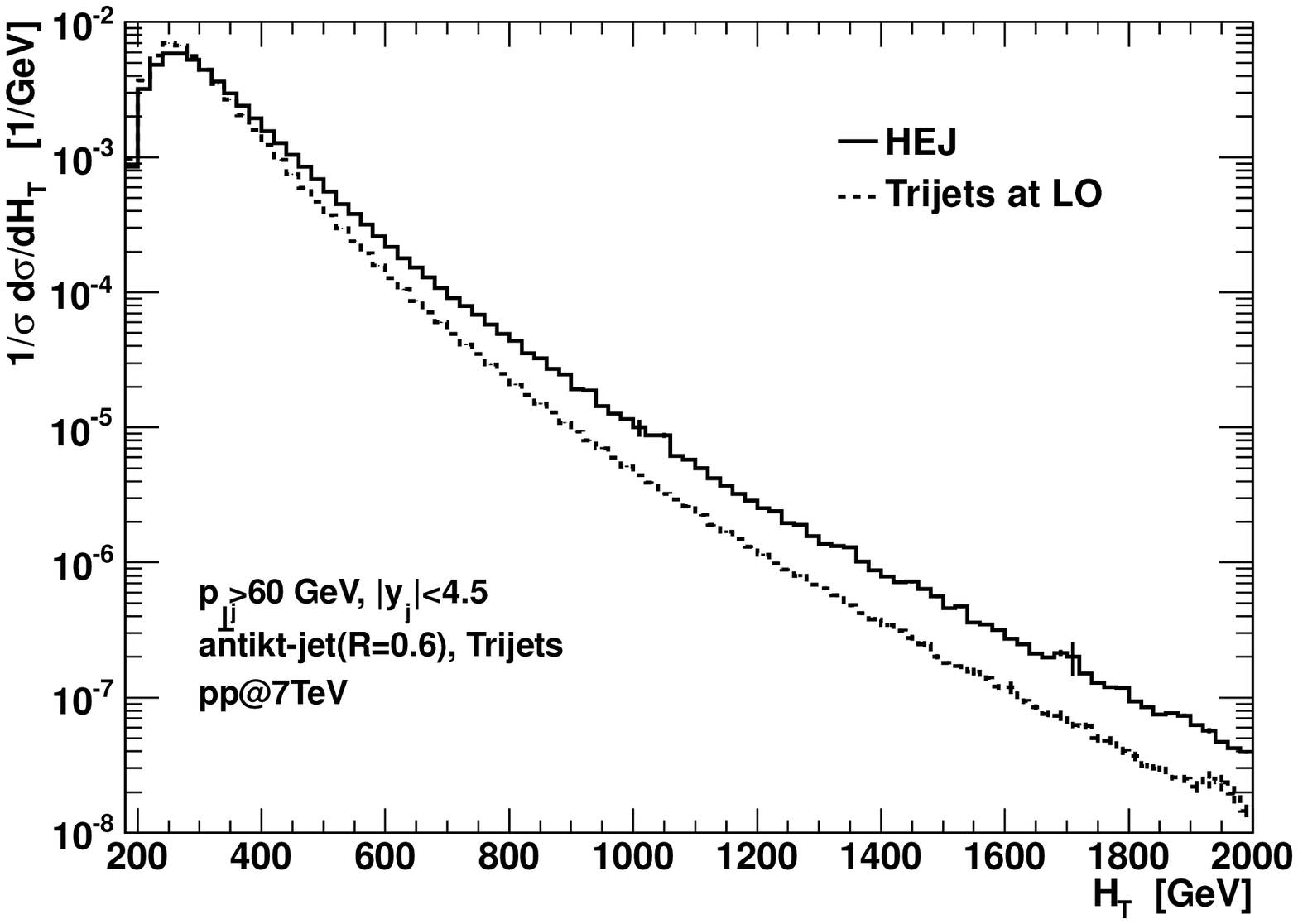}
  \epsfig{width=.49\textwidth,file=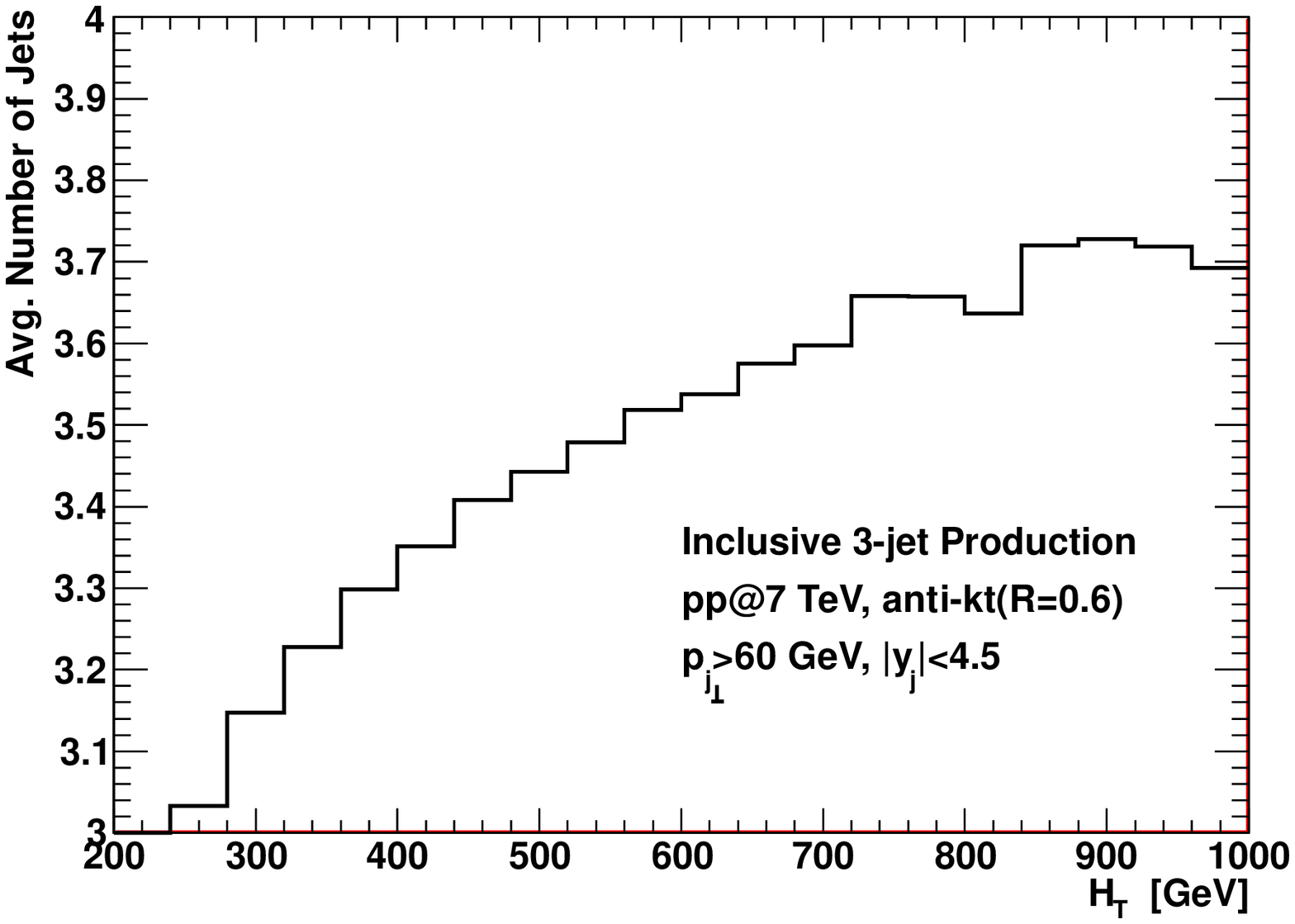}
  \epsfig{width=.49\textwidth,file=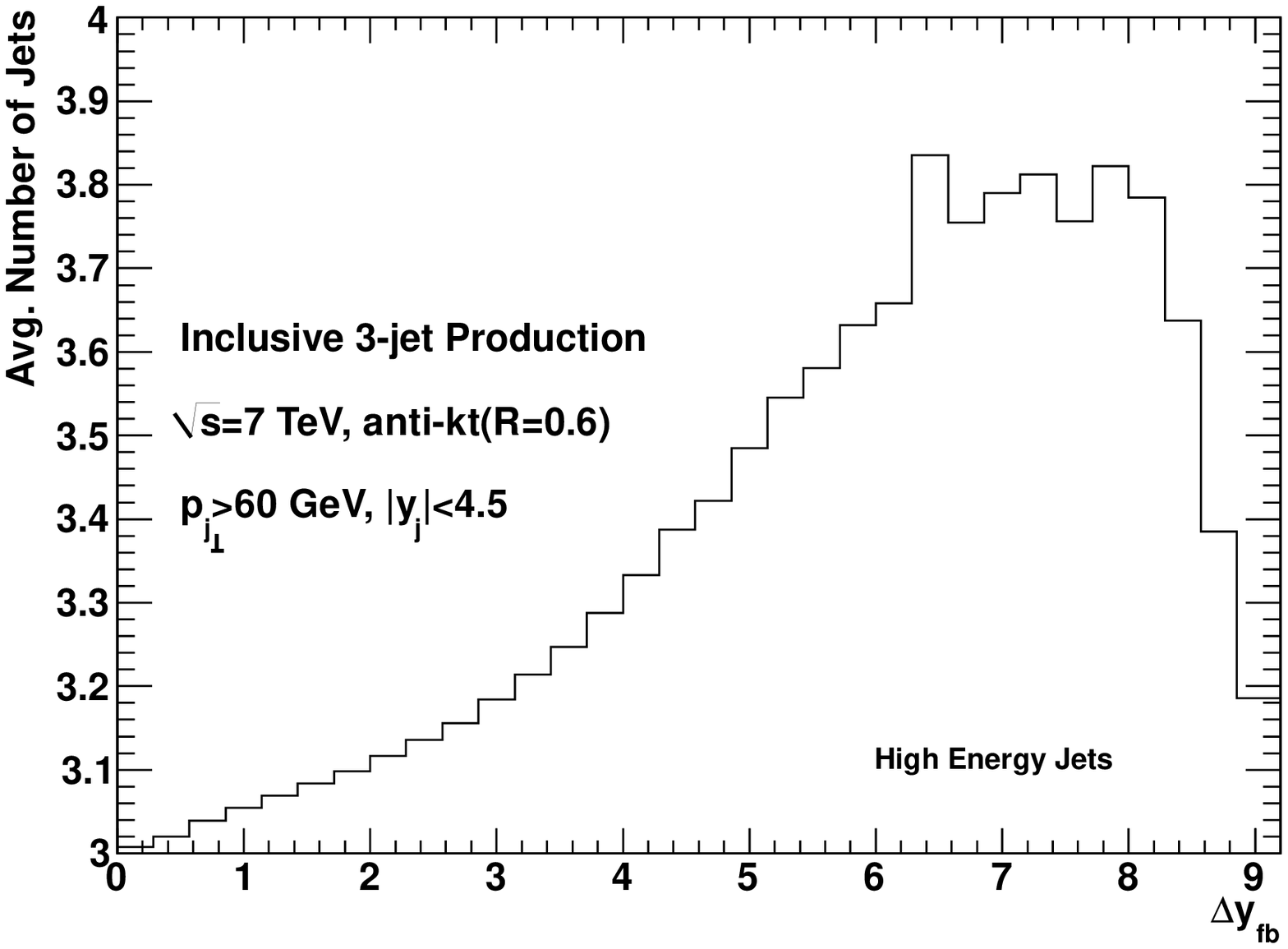}
  \caption{The differential 3-jet cross section with respect to $H_T$ (top left),
    and the normalised spectrum (top right). The radiative corrections
    implemented in \emph{High Energy Jets} enhance the high-$H_T$-tail
    significantly. The bottom plot is of the average number of hard jets
    (transverse momentum above 60GeV) in the events as a function of
    $H_T$. Hard radiative corrections, as those included in HEJ, are clearly
    important in the description of events with large $H_T$.}
  \label{fig:3jht}
\end{figure}
\begin{figure}
  \centering
  \epsfig{width=.49\textwidth,file=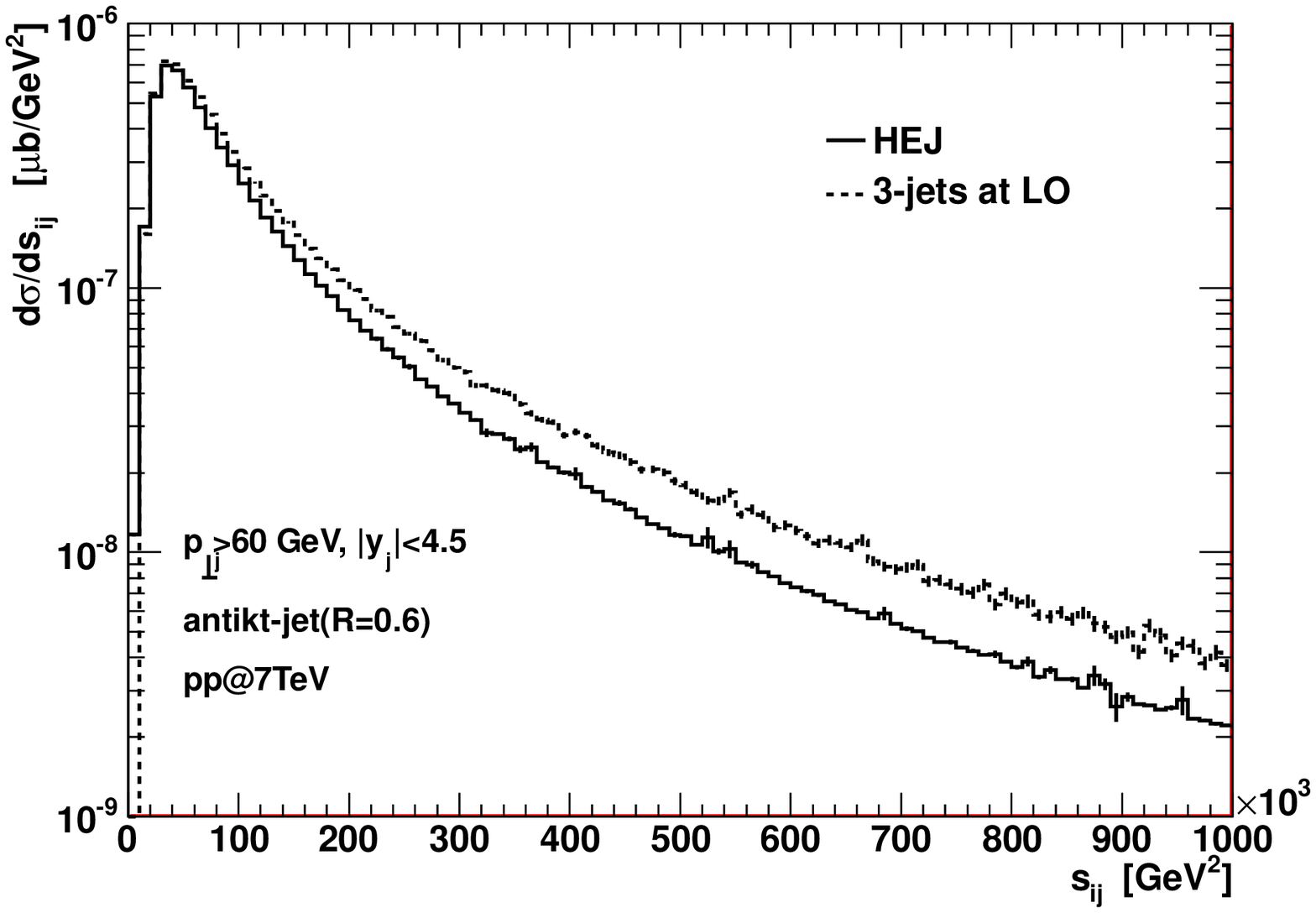}
  \epsfig{width=.49\textwidth,file=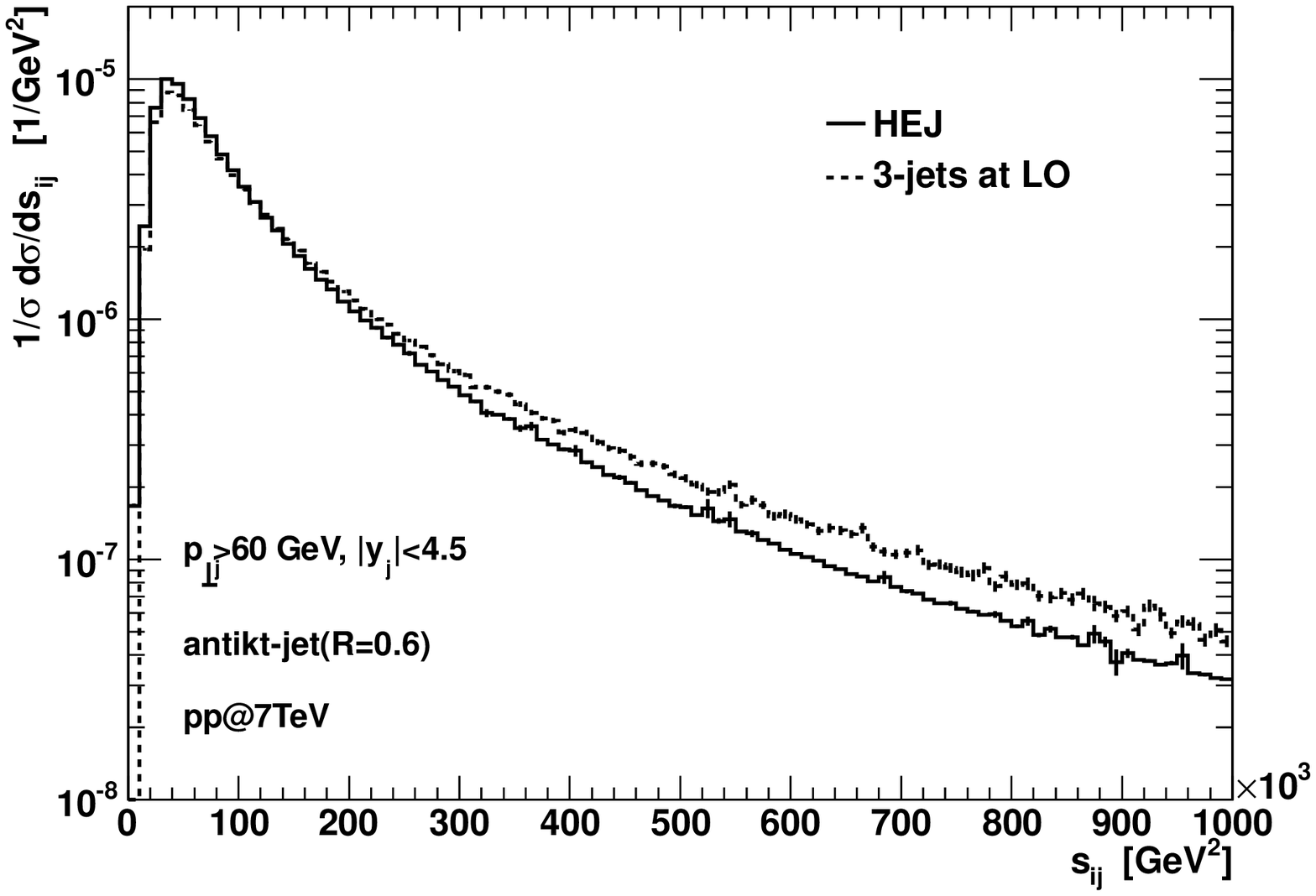}
  \epsfig{width=.49\textwidth,file=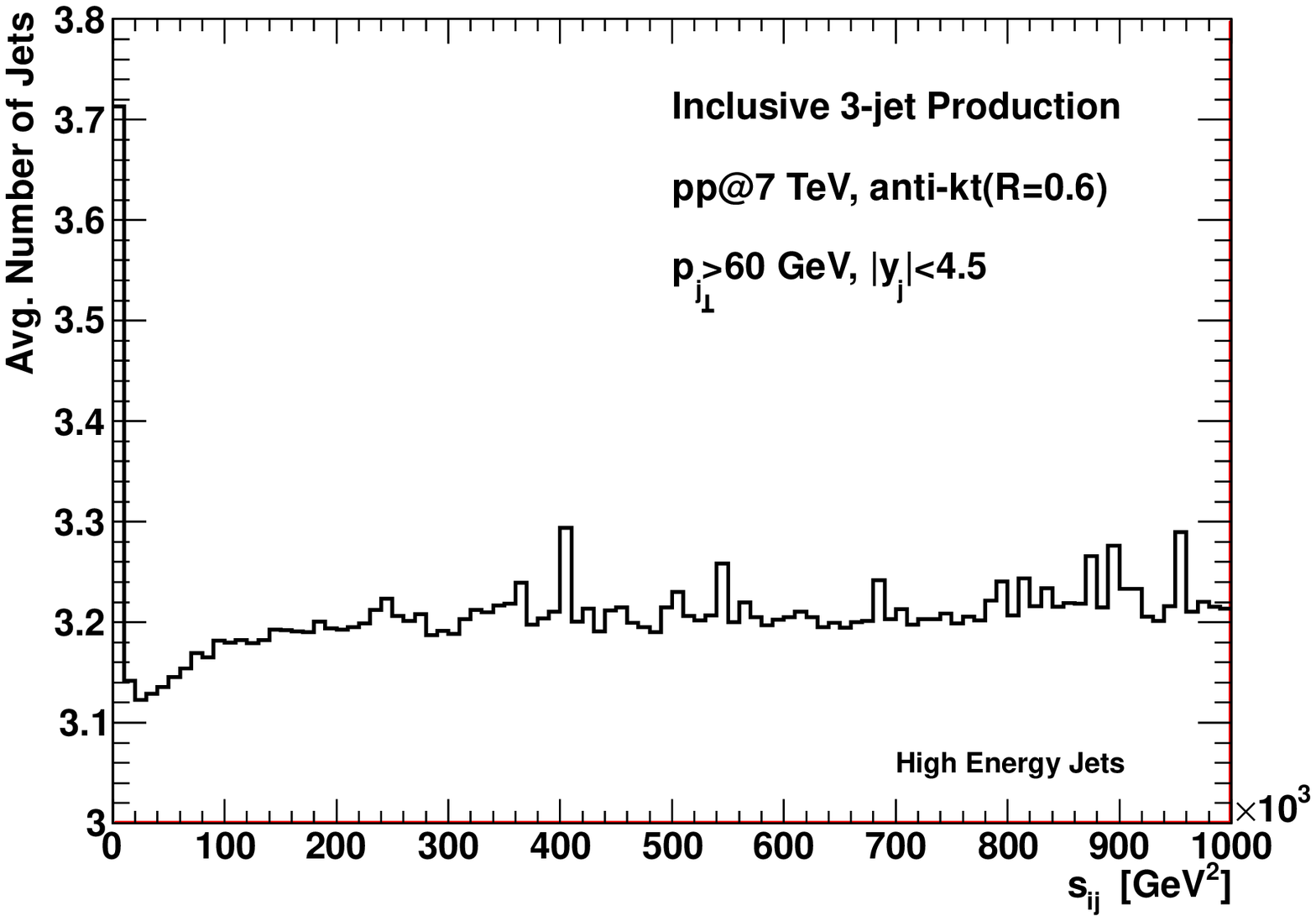}
  \caption{The differential 3-jet cross section with respect to the square of
    the invariant mass $s_{ij}$ between the two hardest jets (top left), and
    the normalised spectrum (top right). The bottom plot is of the average
    number of hard jets (transverse momentum above 60GeV) in the events as a
    function of $s_{ij}$. Please see text for discussion.}
  \label{fig:3jsij}
\end{figure}
In this section we will briefly present the distributions discussed
previously, but this time for events with at least three hard jets. Many of
the features identified in inclusive dijet production, like e.g.~the strong
correlation between the average number of jets and the rapidity difference
between the most forward and most backward hard jet are found also for 3-jet
production.

On the top of Fig.~\ref{fig:3jht} we compare the results for the
$H_T$-distribution in 3-jet production as obtained in leading order and in
\emph{High Energy Jets}. Top left is the distribution in absolute numbers, on
the top right the distribution is divided by the total 3-jet cross
section. Just as in the dijet case, the higher order corrections implemented
in HEJ hardens the spectrum in $H_T$. The average number of hard jets
vs.~$H_T$ is shown on the lower left of Fig.~\ref{fig:3jht}. The distribution
rises to 3.6 at $H_T=800$~GeV and then drops off slightly for increasing
$H_T$. We note that the in the trijet case, the average number of jets rises
0.6 units above the minimum required, whereas in the dijet case it rises a
full 1.2 units. Both cases represent of course large corrections to the
simplistic tree-level point of view. The lower right plot on
Fig.~\ref{fig:3jht} is of the average number of jets vs.~the rapidity span
between the most forward/backward hard jets. It rises from 3 to roughly 3.8
at rapidity differences between 7 and 8, before dropping back down towards 3,
again because the increase in $x$ necessary for additional radiation
leads to a pdf suppression. An increase in the centre-of-mass energy of the
proton-proton collision will obviously lead to a further increase in the number
of hard jets produced.

In Fig.~\ref{fig:3jsij} we study the distributions in the invariant mass
between the two hardest jets in the event. Similarly to the dijet-case, we
find that the results from \emph{HEJ} are suppressed at large $s_{ij}$
compared to the 3-jet LO estimate. The bottom plot on Fig.~\ref{fig:3jsij} is
of the average number of jets in inclusive 3-jet events as a function of the
invariant mass between the two hardest jets. The distribution is very
strongly peaked at small $s_{ij}$ for exactly the same reason as the dijet
case: the correction from additional jet production allows for two central
(and thus generally slightly harder in transverse momentum) jets, which can
form a system of small invariant mass. In fact, the average number of jets in
inclusive 3-jet production at small $s_{ij}$ is almost identical to the
average number of jets in inclusive dijet production at small $s_{ij}$.

\subsection{Gaps Between Di-Jets}
\label{sec:gaps-between-jets}
It is possible to construct several observables which are sensitive to additional
radiation from the dijet-system, and thus can be used as a direct test of the description
arrived at using various descriptions like e.g.~fixed order, shower,
Cascade\cite{Jung:2000hk,Jung:2010si} and \emph{HEJ}. There is a small challenge in
defining quantities which are stable within each perturbative framework. We have already
discussed that the NLO calculation for dijet production is unstable in a setup of equal
transverse momentum cuts on the two jets\cite{Frixione:1997ks,Andersen:2001kt}, but that
it can be stabilised by requiring e.g.~cuts of 65~GeV and 60~GeV on the hardest and
next-to-hardest jet. Some dijet observables will then be calculable in several
frameworks. A well-studied example of an ``inclusive'' dijet observable is the average of
$\cos(\pi-\phi_{jj})$ vs.~$\Delta y_{fb}$. At LO, the two jets are obviously back-to-back,
and $\cos(\pi-\phi_{jj})=1$. The benefit of this observable is that it is completely
inclusive in the radiation between the two jets (this inclusiveness allows for studies
also within (semi-)analytic approaches of BFKL\cite{Colferai:2010wu,Kepka:2010hu}) - any
emission will cause a decorrelation, whether or not it is identified as a separate jet. An
experimental study of this quantity could serve as a strong test of the description of
higher order corrections.

Instead of studying the effect of additional radiation through its impact on
the jets extremal in rapidity, one can study the radiation in-between the
dijets directly. The Atlas collaboration have published a note on such a
study in early data from the LHC\cite{Atlas:2010xx}. They present data for
the so-called ``gap fraction'', defined as the fraction of dijet events with
no additional hard jets between the two (in rapidity). We have already seen
(e.g.~Fig.~\ref{fig:scalechoice}) that the average number of hard jets in
dijet events increases with the rapidity difference between the
forward/backward jet, and this should obviously be reflected in the ``gap
fraction''. These early studies also serve to guide jet veto studies for
Higgs boson production in association with
dijets\cite{Barger:1995zq,Binoth:2010ra}. For these studies, it is of
interest to use a small transverse scale for the vetoing of further jets. The
Atlas study defined jets using the anti-$k_t$ algorithm, with $R=0.6$, and a
transverse scale of 30~GeV. In order to ensure a sufficiently small dijet-rate (and thus an
acceptable scaling factor for the trigger), a harder scale was required. We
will here concentrate on the part of the study, where the average transverse
momentum of the two jets extremal in rapidity was required to be above 60~GeV.
The cuts used are then:
\begin{align}
  \label{eq:dijetatlascuts}
  p_{j_\perp}>30 \mathrm{GeV}\quad \bar{p}_{\perp}>60\mathrm{GeV}\quad |y_j| < 4.5\quad
  \mathrm{anti-kt}, R=0.6.
\end{align}
where $\bar{p}_\perp$ is the average transverse momentum of the most
forward/backward jets. 

\begin{figure}[btp]
  \centering
  \epsfig{width=0.49\textwidth,file=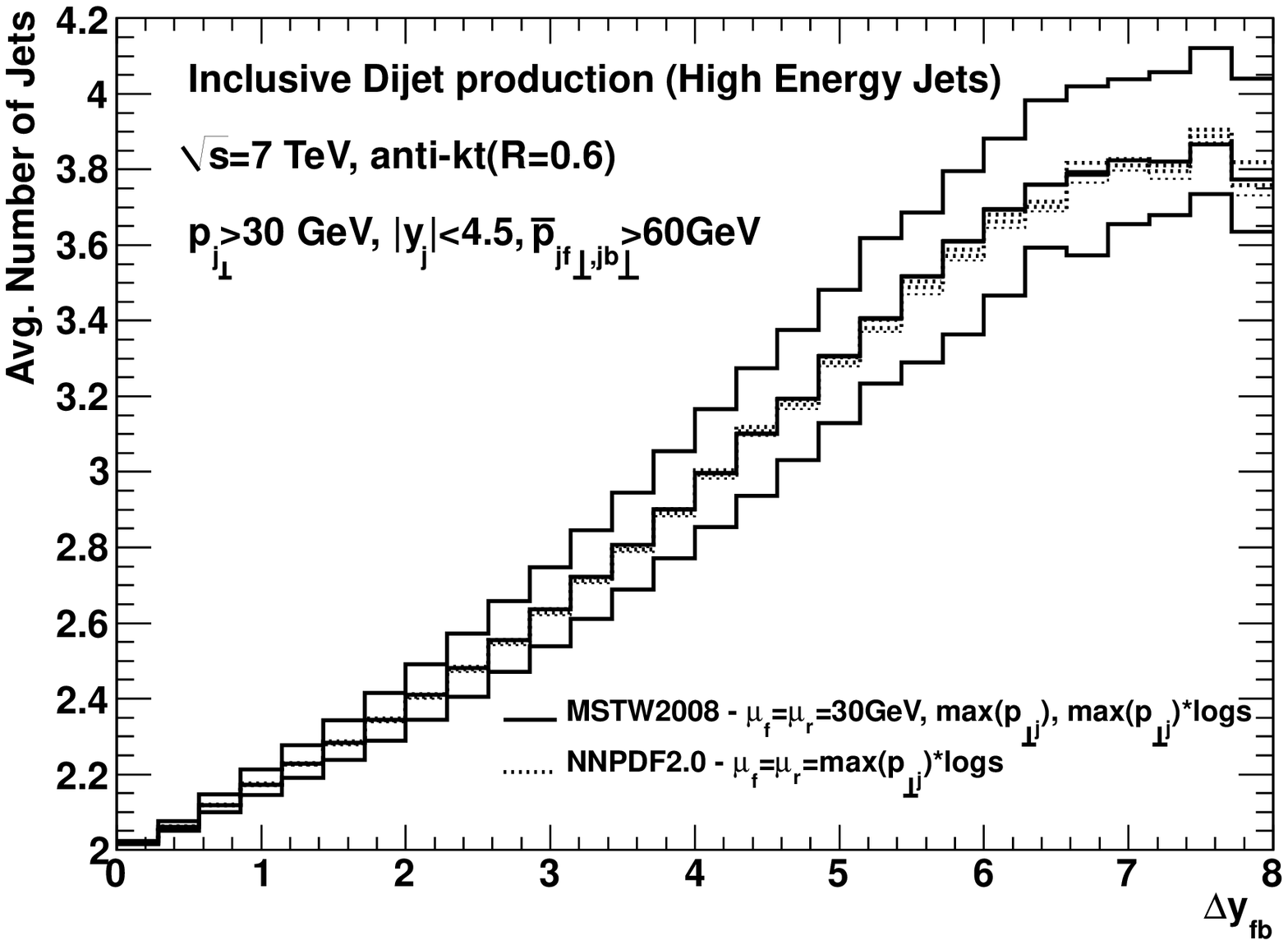}
  \epsfig{width=0.49\textwidth,file=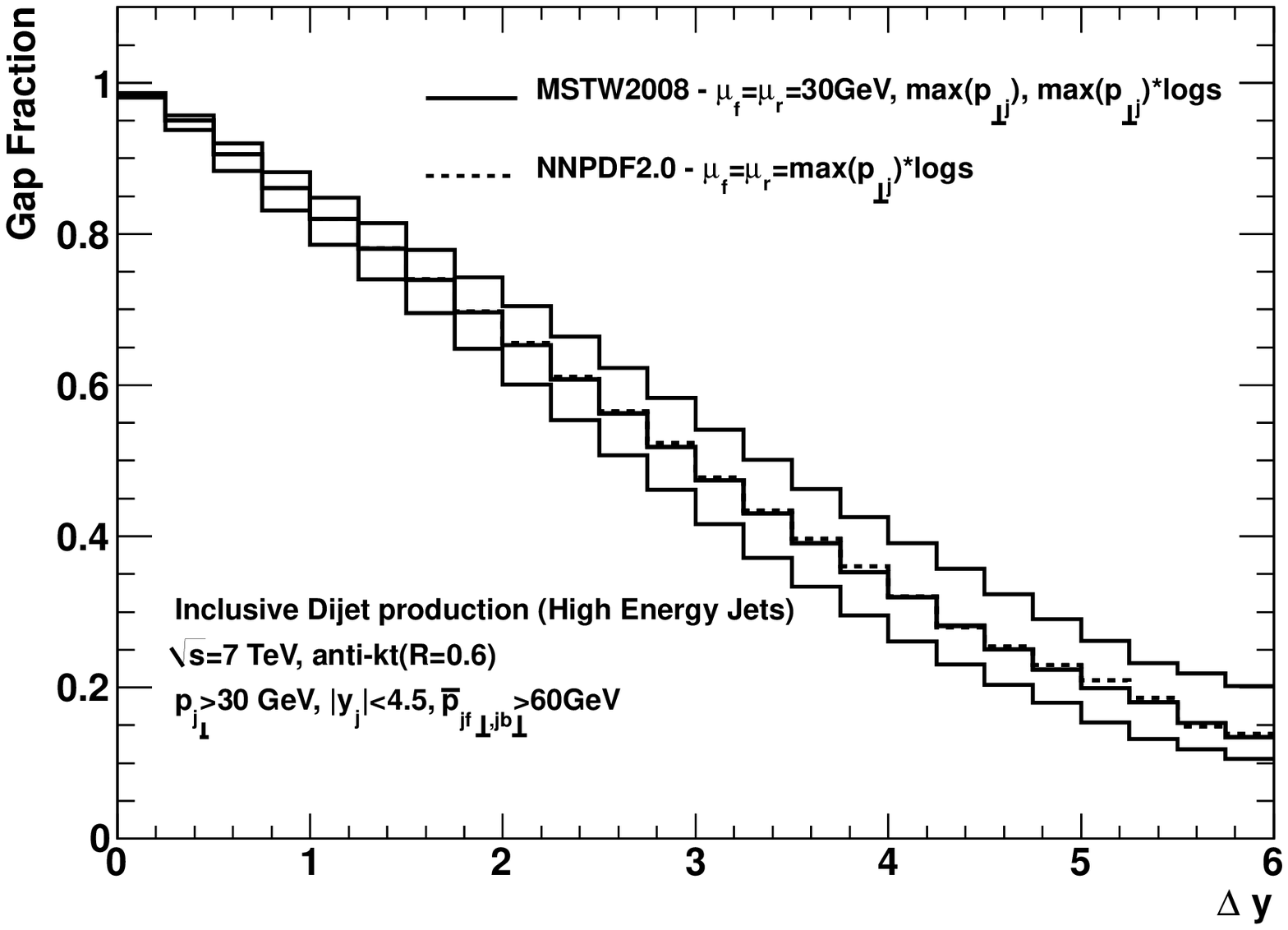}
  \caption{The average number of jets and gap fraction vs. the rapidity difference $\Delta
    y_{fb}$ between the most forward and most backward jets.  The upper and lower solid
    lines are for scale choices of 30~GeV and $\max(p_{\perp j})$ respectively.  The
    central solid lines are for a scale choice of $\max(p_{\perp j})$ plus the logarithms
    of Sec.~\ref{sec:running-coupling} for MSTW2008~\cite{Martin:2009iq} uncertainty pdf
    sets.  The difference between the sets is barely observable.  Also shown in dotted and
    dashed lines are the results with the uncertainty sets from
    NNPDF2.0~\cite{Ball:2010de}.}
  \label{fig:atlasgaps}
\end{figure}
Figure~\ref{fig:atlasgaps} presents the prediction for both the average
number of hard jets (with a transverse momentum larger than 30~GeV) and the
gap fraction obtained using \emph{HEJ}, within the cuts in
eq.~\eqref{eq:dijetatlascuts}. We have also indicated the variation in both
quantities between a scale choice of 30~GeV, $\max(p_{\perp j})$ and of
$\max(p_{\perp j})$ including the logarithmic corrections discussed in
Sec.~\ref{sec:running-coupling}, all using the pdfs included in
MSTW2008\cite{Martin:2009iq}. For the last, ``central'' scale choice we also
present the results obtained by using NNPDF2.0\cite{Ball:2010de}, including
the full envelope of the 100 uncertainty pdfs. The uncertainty induced by the
pdfs on these quantities is completely negligible (they begin to play a role
at $\Delta y_{fb}>8$). The uncertainty estimate induced by a variation in the
renormalisation and factorisation scale between $30$~GeV (the minimum
transverse scale for jets) and $\max(p_{\perp j})$ is increasing for
increasing rapidity spans, and amounts to a variation between 3.6 and 4.0 in
the average number of jets (with a transverse momentum larger than 30~GeV) for
a rapidity span of 7. These results are marked by the outer solid lines (on
both plots). The central solid line is obtained by choosing the
renormalisation scale $\max(p_{\perp j})$, but including the logarithms as
discussed in Sec.~\ref{sec:running-coupling}. The results obtained by
choosing the renormalisation scale as 30~GeV and including the logarithms is
almost identical.

The pdf and scale uncertainty of the predictions for the average number of
jets and the gap fraction are sufficiently small that the the ideas and
calculations presented here can be meaningfully confronted with data, once it
has become available.


\section{Summary and Conclusions}
\label{sec:conclusions}
We have discussed the implementation of the framework of \emph{High Energy
  Jets}\cite{Andersen:2009nu,Andersen:2009he} in a flexible Monte Carlo; the
new components discussed in the present paper include 1) the organisation of
the all-order cancellation of IR divergences between real and virtual
corrections (Sec.~\ref{sec:regul-cross-sect}), 2) matching to high
multiplicity tree-level matrix elements (Sec.~\ref{sec:matching}), and 3) the
inclusion of higher order logarithmic terms to stabilise the scale dependence
(Sec.~\ref{sec:running-coupling}).

In Sec.~\ref{sec:dijet-event-results} we studied the impact of the
perturbative corrections included in \emph{High Energy Jets} on a number of
dijet and trijet distributions. We find that compared to LO, the distribution
is harder in both the transverse momentum and in $H_T$ (the scalar sum of
transverse jet momenta), while the invariant-mass distribution between the
two hardest jets is softened. Similar results hold for
trijet-observables. Therefore, the understanding of the radiative corrections
could lead to better methods for suppressing the Standard Model contribution
in new-physics searches.

The effect of hard, perturbative corrections is cleanly displayed by the
average number of hard jets versus the observable in question. This is
particularly true for the rapidity span between the most forward/backward
hard jets, which is a direct measure of the phase space available for hard
radiation.

Finally, in Sec.~\ref{sec:gaps-between-jets}, we presented the prediction
obtained from \emph{High Energy Jets} of an observable sensitive to inter-jet
radiation, which is currently under study by Atlas\cite{Atlas:2010xx}. We
find that the theoretical uncertainty on the quantities studied is
dominated by the scale choice, while the variation induced by pdf
uncertainties is completely negligible.  

The generator \emph{High Energy Jets} is available at
\texttt{http://cern.ch/hej}.  

\section{Acknowledgements}
\label{sec:acknowledgements}

We thank Leif L\"onnblad and Jeff Forshaw for interesting discussions.


\appendix
\section{The Building Blocks for the Regularised \emph{High Energy Jets}-Cross Sections}
\label{cha:build-blocks-regul}
We define here in one place the necessary building blocks to construct an amplitude in the
HEJ framework:
\begin{align}
  \label{eq:MHEJagain}
  \begin{split}
    \overline{\left|\mathcal{M}_{\rm HEJ}^\mathrm{reg}(\{ p_i\})\right|}^2 = \ &\frac 1 {4\
       (\Nc^2-1)}\ \left\|S_{f_1 f_2\to f_1 f_2}\right\|^2\\
     &\cdot\ \left(g^2\ K_{f_1}\ \frac 1 {t_1}\right) \cdot\ \left(g^2\ K_{f_2}\ \frac 1
       {t_{n-1}}\right)\\
     & \cdot \prod_{i=1}^{n-2} \left( {g^2 C_A}\
       \left(\frac {-1}{t_it_{i+1}} V^\mu(q_i,q_{i+1})V_\mu(q_i,q_{i+1}) -
         \frac{4}{\mathbf{p}_i^2}\ \theta\left(\mathbf{p}_i^2<\lambda^ 2\right)\right)\right)\\
     & \cdot \prod_{j=1}^{n-1} \exp\left[\omega^0(q_j,\lambda)(y_{j-1}-y_j)\right].
   \end{split}
\end{align}
Our momentum convention will be that $p_{A}$ and $p_B$ represent the momenta of the
forward and backward moving initial partons respectively.  The outgoing momenta of all
quarks and gluons are then numbered in decreasing rapidity so $p_1$ is the most forward
etc.  We then define $q_i$ to be the momenta which correspond to the $t$-channel momenta
in the effective $t$-channel exchange picture, that is
\begin{align}
  \label{eq:defqi}
  q_1=p_A-p_1, \qquad q_i=q_{i-1}-p_i\quad 2\le i \le n-1.
\end{align}

The current pieces, $\left\|S_{f_1 f_2\to f_1 f_2}\right\|^2$, indicate the square of pure
current-current scattering. For quarks this is
\begin{align}
  \label{eq:SdefQ}
    \left\|S_{qQ\to qQ}\right\|^2 = \left|j_{a1}^{-}\cdot j_{bn}^{-}\right|^2 +
    \left|j_{a1}^{-} \cdot j_{bn}^{+}\right|^2 + \left|j_{a1}^{+}\cdot
      j_{bn}^{-}\right|^2 + \left|j_{a1}^{+}\cdot j_{bn}^{+} \right|^2.
\end{align}
Anti-quarks are treated in the same way with $j_{a1}^- \to \bar j_{a1}^- = \bar v_a^-
\gamma^\mu v_1^-$.  For gluons it is more complicated as there is an overall factor for
helicity conserving channels (see eq.~\eqref{eq:sqcol}) and also flipped-helicity
contributions weighted with different factors~\cite{Andersen:2009he}:
\begin{align}
  \label{eq:Sdefg}
  \begin{split}
    \left\|S_{qg\to qg}\right\|^2 =& \left( \frac12\ \frac{1+z^2}z\ \left(1 -\frac 1
        {C_A^2}\right)+\frac{1}{C_A^2} \right)\times \\ & \left( \left|j_{a1}^{-}\cdot
        j_{bn}^{-}\right|^2 + \left|j_{a1}^{-} \cdot j_{bn}^{+}\right|^2 +
      \left|j_{a1}^{+}\cdot j_{bn}^{-}\right|^2 + \left|j_{a1}^{+}\cdot j_{bn}^{+}
      \right|^2 \right)\\
    &+ \left| S_{q^- g^-\to q^- g^+} \right|^2 + \left| S_{q^+ g^+ \to q^+ g^-} \right|^2
    + \left| S_{q^-g^+\to q^- g^-} \right|^2 + \left| S_{q^+ g^- \to q^+ g^+} \right|^2.
  \end{split}
\end{align}
For this example of a backward-moving parton $z=p_n^-/p_b^-$.  The result for a
forward-moving gluon is the same with $p_a \leftrightarrow p_b$, $p_1 \leftrightarrow
p_n$ and $z=p_1^+/p_a^+$.  The helicity-flipped contributions are given by
\begin{align}
  \label{eq:flips}
  \begin{split}
    \left| S_{q^- g^-\to q^- g^+} \right|^2 =& \frac12 \left( 1- \frac1{C_A^2}\right)
    \left( z \left| j_{a1}^- \cdot j_{b2}^+ \right|^2 + \frac1z \left| j_{a1}^- \cdot j_{bn}^-
      \right|^2 \right) + \left| (p_b+p_n)\cdot j_{a1}^- \right|^2 \\ &+\sqrt{z}
    \ \Re \left[ \frac{p_{n\perp}^*}{|p_{n\perp}|}\
      \cdot\ (p_n+p_b).j_{a1}^-\ \cdot\ (j^+_{bn} \cdot j^-_{a1})^*\right]\\
    &+\sqrt{\frac1z}\ \Re\left[\frac{p_{n\perp}}{|p_{n\perp}|}\
      \cdot \
      (p_n+p_b).j^-_{a1}\ \cdot \ (j^-_{bn} \cdot j^-_{a1})^*\right]\\
    &+\Re\left[\frac{p_{n\perp}}{p_{n\perp}^*} j^+_{bn} \cdot j^-_{a1}\ \cdot\ (j^-_{bn}
      \cdot j^-_{a1})^* \right] \quad = \left| S_{q^+ g^+ \to q^+ g^-} \right|^2
  \end{split}
\end{align}
and
\begin{align}
  \label{eq:flips2}
  \begin{split}
    \left| S_{q^- g^+\to q^- g^-} \right|^2 =& \frac12 \left( 1- \frac1{C_A^2}\right)
    \left( z \left| j_{a1}^- \cdot j_{b2}^- \right|^2 + \frac1z \left| j_{a1}^- \cdot j_{bn}^+
      \right|^2 \right) + \left| (p_b+p_n)\cdot j_{a1}^- \right|^2 \\ &+\sqrt{z}
    \ \Re \left[ \frac{p_{n\perp}}{|p_{n\perp}|}\
      \cdot\ (p_n+p_b).j_{a1}^-\ \cdot\ (j^-_{bn} \cdot j^-_{a1})^*\right]\\
    &+\sqrt{\frac1z}\ \Re\left[\frac{p_{n\perp}^*}{|p_{n\perp}|}\
      \cdot \
      (p_n+p_b).j^-_{a1}\ \cdot \ (j^+_{bn} \cdot j^-_{a1})^*\right]\\
    &+\Re\left[\frac{p_{n\perp}^*}{p_{n\perp}} j^-_{bn} \cdot j^-_{a1}\ \cdot\ (j^+_{bn}
      \cdot j^-_{a1})^* \right] \quad = \left| S_{q^+ g^- \to q^+ g^+} \right|^2.
  \end{split}
\end{align}
Gluon-gluon scattering is the natural generalisation of what has gone before.  There are
now two relevant ratios $z_1=p_1^+/p_a^+$ and $z_2=p_2^-/p_b^-$. Then we define
\begin{align}
  \label{eq:Sggoverall}
  \begin{split}
    \left\| S_{gg\to gg} \right\|^2 =& \left( \frac12\ \frac{1+z_1^2}{z_1}\ \left(1 -\frac 1
        {C_A^2}\right)+\frac{1}{C_A^2} \right)\left( \frac12\ \frac{1+z_2^2}{z_2}\ \left(1
        -\frac 1 
        {C_A^2}\right)+\frac{1}{C_A^2} \right)\times \\ 
    & \left( \left|j_{a1}^{-}\cdot j_{bn}^{-}\right|^2 + \left|j_{a1}^{-} \cdot
        j_{bn}^{+}\right|^2 + \left|j_{a1}^{+}\cdot j_{bn}^{-}\right|^2 +
      \left|j_{a1}^{+}\cdot j_{bn}^{+} \right|^2 \right)\\
    &+ \left( \frac12\ \frac{1+z_1^2}{z_1}\ \left(1 -\frac 1
        {C_A^2}\right)+\frac{1}{C_A^2} \right) \times \\
    & \left( \left| S_{g^- g^-\to g^- g^+} \right|^2 + \left| S_{g^+ g^+ \to g^+ g^-} \right|^2
      + \left| S_{g^-g^+\to g^- g^-} \right|^2 + \left| S_{g^+ g^- \to g^+ g^+} \right|^2
    \right)\\
    &+ \left( \frac12\ \frac{1+z_2^2}{z_2}\ \left(1
        -\frac 1 
        {C_A^2}\right)+\frac{1}{C_A^2} \right) \times \\
    & \left( \left| S_{g^- g^-\to g^+ g^-} \right|^2 + \left| S_{g^+ g^+ \to g^- g^+} \right|^2
    + \left| S_{g^+g^-\to g^- g^-} \right|^2 + \left| S_{g^- g^+ \to g^+ g^+} \right|^2\right).
  \end{split}
\end{align}
As the necessary factor has already been included explicitly in eq.~\eqref{eq:Sggoverall},
the helicity-specific terms like $\left| S_{g^- g^- \to g^- g^+} \right|^2$ are equal to
the corresponding quark-gluon ones if the helicity-conserving gluon was replaced by a
quark, e.g.
\begin{align}
  \label{eq:ggex}
  \left| S_{g^- g^- \to g^- g^+} \right|^2 = \left| S_{q^- g^- \to q^- g^+} \right|^2,
  \quad \left| S_{g^+ g^+ \to g^- g^+} \right|^2 = \left| S_{g^+ q^+ \to g^- q^+} \right|^2.
\end{align}
We do not include the contributions where neither gluon conserves helicity (although this
could be generalised) because the contribution is so insignificant.

Returning to the remaining pieces of eq.~\eqref{eq:MHEJagain}, the factors $K_{f_1}$ are
straight-forward and inspired by the exact high-energy limit: $K_q=\Cf$ for a quark of
any flavour and $K_g=\Ca$ for gluons.

The emission vertices were given in eq.~\eqref{eq:GenEmissionV}:
\begin{align}
  \label{eq:GenEmissionVapp}
  \begin{split}
  V^\rho(q_i,q_{i+1},p_A,p_B,p_1,p_n)=&-(q_i+q_{i+1})^\rho \\
  &+ \frac{p_A^\rho}{2} \left( \frac{q_i^2}{p_{i+1}\cdot p_A} +
  \frac{p_{i+1}\cdot p_B}{p_A\cdot p_B} + \frac{p_{i+1}\cdot p_n}{p_A\cdot p_n}\right) +
p_A \leftrightarrow p_1 \\ 
  &- \frac{p_B^\rho}{2} \left( \frac{q_{i+1}^2}{p_{i+1} \cdot p_B} + \frac{p_{i+1}\cdot
      p_A}{p_B\cdot p_A} + \frac{p_{i+1}\cdot p_1}{p_B\cdot p_1} \right) - p_B
  \leftrightarrow p_n.
  \end{split}
\end{align}

The combination of the virtual corrections in Section \ref{sec:all-orders-virtual} and the
regularisation in Section~\ref{sec:regul-cross-sect} give the following factor in the
exponential in eq.~\eqref{eq:MHEJagain}:
\begin{align}
  \label{eq:omega0}
       \omega^0(q_j,\lambda)=\ &-\frac{\alpha_s N_C}{\pi} \log\frac{{\bf q}_j^2}{\lambda^2}
\end{align}
where the bold indicates that it is the sum of the square of the transverse components
which are included in the log.

\boldmath
\section{Variations of the regularisation parameter $\lambda$}
\label{sec:vari-regul-param}
\unboldmath 
In this appendix, we show a few results to demonstrate that our conclusions are not
sensitive to the chosen value of the regularisation parameter $\lambda$.  This is the
scale above which radiation is considered to be a real emission. The regularisation
procedure is described in full in Section~\ref{sec:regul-cross-sect}.  We again use the
cuts defined in eq.~\eqref{eq:dijetcuts} throughout.

Table~\ref{tab:lambdacxs} shows the exclusive $n$-jet rates for $n=2,3,4$, for values of
$\lambda$ from 0.2--2~GeV.  We can see that the changes in $\lambda$ do not have a large
effect, particularly for the $2j$ and $3j$ rates. 
\begin{table}[hbt]
  \centering
  \begin{tabular}{|c|c|c|c|}
    \hline $\lambda$ (GeV)& $\sigma(2j)$ ($\mu$b) & $\sigma(3j)$ ($\mu$b) & $\sigma(4j)$
    ($\mu$b) \\ \hline
    0.2 & 1.58 & 5.90 E-2 & 9.6$\pm$0.1 E-3\\
    0.5 & 1.58 & 5.93 E-2 & 10.1$\pm$0.1 E-3\\
    1.0 & 1.59 & 5.95 E-2 & 9.7$\pm$0.2 E-3\\
    2.0 & 1.61 & 5.99 E-2 & 9.2$\pm$0.2 E-3\\ \hline
  \end{tabular}
  \caption{Exclusive $n$-jet cross sections for different values of the regularisation
    parameter $\lambda$.  The errors shown are statistical -- they are not shown for the
    $2j$ and $3j$ rates because they are smaller than the last quoted digit.}
  \label{tab:lambdacxs}
\end{table}

\begin{figure}[btp]
  \centering
  \epsfig{file=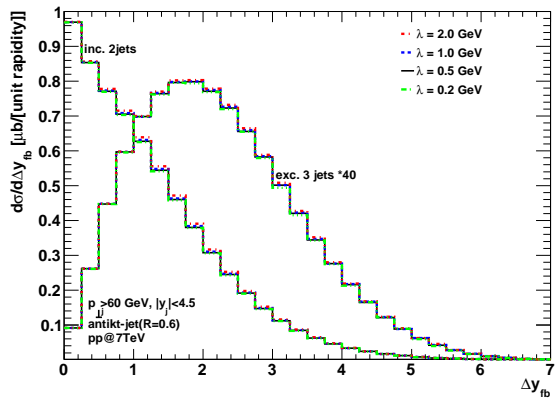,width=0.5\textwidth}
  \caption{This plot shows the $\Delta y_{fb}$ distribution, for different values of the
    regularisation parameter $\lambda$ for the inclusive 2-jet sample and the exclusive
    3-jet sample (times 40).  The differences are small and we choose to use a default
    value of 0.5~GeV.}
  \label{fig:ydiflambda}
\end{figure}

Figure~\ref{fig:ydiflambda} shows the distribution of the rapidity difference between most
forward and most backward jet, $\Delta y_{fb}$, for different values of $\lambda$ both for
the inclusive 2-jet sample and the exclusive 3-jet sample.  The differences are very
small.  We use $\lambda=0.5$~GeV as the default.

In the HEJ framework, the number of quarks and gluons is treated as a variable and
contributions are summed over $n$ from 2 to $\infty$, see eq.~\eqref{eq:dijetreg}.  In
practice, there is an upper cut-off on the value of $n$.  This has a very weak dependence
on $\lambda$ as it stands to reason that the lower the cut-off on resolved emissions, the
more of them you need to consider to get the same results.  For $\lambda=0.5$~GeV, we use
$n_{\rm max}=22$, and find no observable difference in physical results by varying around
this value.

\section{Stability of Equal Cut in the Transverse Momentum of Dijets}
\label{sec:stability-equal-cut}

\begin{figure}[btp]
  \centering
  \epsfig{file=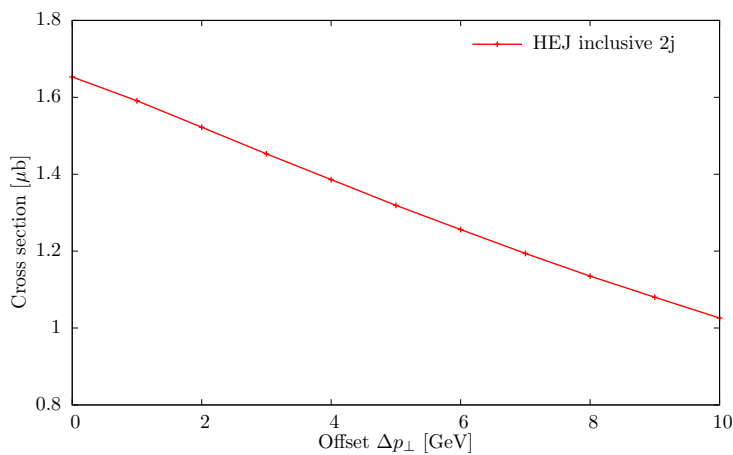,width=0.6\textwidth}
  \caption{The inclusive dijet cross section from \emph{HEJ} as a function of the
    offset-parameter $\Delta p_\perp$.} 
  \label{fig:offset}
\end{figure}

As discussed in Section~\ref{sec:generic-multi-jet}, NLO calculations suffer from an
instability when the transverse momentum cuts on the two jets are
symmetric~\cite{Frixione:1997ks}.  The same effect however is not seen in other
calculations which are not terminated at NLO~\cite{Alioli:2010xa,Andersen:2001kt}.  In
line with this, \emph{HEJ} does not see an instability here either.  We illustrate this by
implementing an offset-parameter, $\Delta p_\perp$, and requiring a difference in the
transverse momenta of the two jets of at least this value.  Figure~\ref{fig:offset} shows
the dijet cross section as a function of $\Delta p_\perp$.  It continues to increase
linearly as $\Delta p_\perp$ decreases to zero and shows no sign of the turnover observed
at NLO.

\bibliographystyle{JHEP}
\bibliography{papers}

\end{document}